\documentclass[12pt,aps,superscriptaddress,prl]{revtex4-2}

\usepackage{graphicx}  
\usepackage{dcolumn}   
\usepackage{bm}        
\usepackage{amssymb}   
\usepackage{amsmath}   
\usepackage{amsthm}    
\usepackage{xcolor}    
\usepackage{tikz}      
\usepackage{ifthen}    
\usepackage{multirow}  
\usepackage{subfigure}
\usepackage{comment}
\usepackage{bbold}
\usepackage{xspace}
\usepackage[unicode=true,
  linktocpage,
  linkbordercolor={0.5 0.5 1},
  citebordercolor={0.5 1 0.5},
  linkcolor=blue]{hyperref}
\newcommand{\bfs}{BaFe$_2$As$_2$\xspace}
\newcommand{\aog}{A$_{1g}$\xspace}
\newcommand{\bog}{B$_{1g}$\xspace}
\newcommand{\btg}{B$_{2g}$\xspace}
\newcommand{\Tc}{$T_c$\xspace}
\newcommand{\TS}{$T_S$\xspace}

\newcounter{para}

\begin{document}
	
\title{Elastocaloric evidence for a multicomponent superconductor stabilized within the nematic state in Ba(Fe$_{1-x}$Co$_x$)$_2$As$_2$}

\author{Sayak Ghosh}
\affiliation{Geballe Laboratory for Advanced Materials, Stanford University, Stanford, CA, USA}
\affiliation{Department of Applied Physics, Stanford University, Stanford, CA, USA}
\author{Matthias S. Ikeda}
\affiliation{Geballe Laboratory for Advanced Materials, Stanford University, Stanford, CA, USA}
\affiliation{Department of Applied Physics, Stanford University, Stanford, CA, USA}
\author{Anzumaan R. Chakraborty}
\affiliation{School of Physics and Astronomy, University of Minnesota, Minneapolis, MN, USA}
\author{Thanapat Worasaran}
\affiliation{Geballe Laboratory for Advanced Materials, Stanford University, Stanford, CA, USA}
\affiliation{Department of Applied Physics, Stanford University, Stanford, CA, USA}
\author{Florian Theuss}
\affiliation{Laboratory of Atomic and Solid State Physics, Cornell University, Ithaca, NY, USA}
\author{Luciano B. Peralta}
\affiliation{Laboratory of Atomic and Solid State Physics, Cornell University, Ithaca, NY, USA}
\affiliation{Department of Physics, Universidad de Los Andes, Bogotá 111711, Colombia}
\author{P. M. Lozano}
\affiliation{Advanced Photon Source, Argonne National Lab, Lemont, IL, USA}
\author{Jong-Woo Kim}
\affiliation{Advanced Photon Source, Argonne National Lab, Lemont, IL, USA}
\author{Philip J. Ryan}
\affiliation{Advanced Photon Source, Argonne National Lab, Lemont, IL, USA}
\author{Linda Ye}
\affiliation{Department of Applied Physics, Stanford University, Stanford, CA, USA}
\author{Aharon Kapitulnik}
\affiliation{Geballe Laboratory for Advanced Materials, Stanford University, Stanford, CA, USA}
\affiliation{Department of Applied Physics, Stanford University, Stanford, CA, USA}
\author{Steven A. Kivelson}
\affiliation{Geballe Laboratory for Advanced Materials, Stanford University, Stanford, CA, USA}
\affiliation{Department of Physics, Stanford University, Stanford, CA, USA}
\author{B.~J.~Ramshaw}
\affiliation{Laboratory of Atomic and Solid State Physics, Cornell University, Ithaca, NY, USA}
\affiliation{Canadian Institute for Advanced Research, Toronto, Ontario, Canada}
\author{Rafael M. Fernandes}
\affiliation{School of Physics and Astronomy, University of Minnesota, Minneapolis, MN, USA}
\author{Ian R. Fisher}
\affiliation{Geballe Laboratory for Advanced Materials, Stanford University, Stanford, CA, USA}
\affiliation{Department of Applied Physics, Stanford University, Stanford, CA, USA}

\maketitle

\textbf{The iron-based high-\Tc superconductors exhibit rich phase diagrams with intertwined phases, including magnetism, nematicity and superconductivity. The superconducting \Tc in many of these materials is maximized in the regime of strong nematic fluctuations, making the role of nematicity in influencing the superconductivity a topic of intense research. Here, we use the AC elastocaloric effect (ECE) to map out the phase diagram of Ba(Fe$_{1-x}$Co$_x$)$_2$As$_2$ near optimal doping. The ECE signature at \Tc on the overdoped side, where superconductivity condenses without any nematic order, is  quantitatively consistent with other thermodynamic probes that indicate a single-component superconducting state. In contrast, on the slightly underdoped side, where superconductivity condenses within the nematic phase, ECE reveals a second thermodynamic transition proximate to and below \Tc. We rule out magnetism and re-entrant tetragonality as the origin of this transition, and find that our observations strongly suggest a phase transition into a multicomponent superconducting state. This implies the existence of a sub-dominant pairing instability that competes strongly with the dominant $s^\pm$ instability. Our results thus motivate a re-examination of the pairing state and its interplay with nematicity in this extensively studied iron-based superconductor, while also demonstrating the power of ECE in uncovering strain-tuned phase diagrams of quantum materials.}

\section{Introduction}

Phases of quantum materials are described by order parameters (OPs) that capture the broken symmetries characteristic of each phase. In a superconductor, the OP is given by the Cooper pair wavefunction, or the superconducting gap. While most superconductors (SCs) exhibit a single-component order parameter—such as $s$-wave in BCS superconductors like lead and aluminum, or $d$-wave in the high-\Tc cuprates—less common are multi-component OPs. The degeneracy of the multiple components can be enforced by symmetry or might be accidental. A classic example of a symmetry-enforced degeneracy is the $p_x+ip_y$ state in a tetragonal lattice \cite{KallinRPP2012}. This was believed to be realized in Sr$_2$RuO$_4$ for a long time, although recent measurements have ruled it out \cite{ChronisterPNAS2021}. In cases of accidental degeneracies, two superconducting instabilities often exist in close energy proximity, as suggested in the recently discovered heavy-fermion superconductor UTe$_2$ \cite{HayesScience2021}. In such scenarios, the dominant pairing mechanism is expected to determine \Tc. However, the sub-dominant instability may produce a second transition if it is sufficiently close in energy or if stabilized by other phases. Understanding how multiple superconducting instabilities may emerge and interact with each other, as well as with other non-SC phases in a material, is central to uncovering the pairing mechanism(s) that exist in a multi-component superconductor \cite{RaghuPRB2012,YuanPRB2023}. In addition, due to their extra degrees of freedom, multicomponent SCs may give rise to emergent effects unattainable in single-component SCs \cite{MilosevicIOP2015,FernandesPRL2021}.

The Fe-based SCs are generally accepted to have an $s^\pm$ order parameter with opposite sign of the gap on the electron and the hole pockets \cite{HirschfeldIOP2011,FernandesRPP2017,Fernandes2022}. The dominant pairing mechanism is believed to be spin fluctuations which promotes both $s^\pm$ and $d$-wave pairing, with $s^\pm$ usually prevailing \cite{GraserPRB2010,MaitiPRL2011}. However, the multiband nature of the materials may stabilize the $d$-wave, or even more exotic OPs---such as $s+is$, $s+d$, or $s+id$---under suitable conditions \cite{KhodasPRL2012,PlattPRB2012}. In K-doped \bfs, for example, the superconducting gap evolves from nodeless $s^\pm$ at low dopings to hosting a complicated pattern of nodes at high dopings \cite{ReidPRL2012,OkazakiSc2012}, which has been interpreted to be evidence for $s+is$ or $d$-wave OP. The question of whether a $d$-wave, or any sub-dominant, instability exists is not simply a matter of detail---it provides crucial insights into how the nematic phase, which is ubiquitous among various Fe-based SCs, and its fluctuations
interact with the superconductivity \cite{FernandesPRL2013,KangPRB2018}. In this work, we study the archetypal Fe-based superconductor: Co-doped \bfs. Near optimal doping, Raman scattering \cite{MuschlerPRB2009,KretzschmarPRL2013} and thermal conductivity \cite{TanatarPRL2010} measurements indeed show evidence for a sub-dominant instability. However, no experimental signature of a second SC transition has been observed in thermodynamic measurements like specific heat. Further, how the sub-dominant instability evolves across optimal doping also remains an open question.
	
While Co-doped \bfs has been studied extensively utilizing chemical doping and through transport measurements under uniaxial strain, thermodynamic measurements under uniaxial strain have been lacking. Previous work by \citet{MeingastPRL2012} indicates that uniaxial strain should tune the material in a similar fashion as Co doping. We employ the AC elastocaloric effect (ECE) to probe the phase diagram near optimal doping under uniaxial stress conditions. The ECE is a highly sensitive, thermodynamic probe of phase transitions, which has been recently employed to study the non-trivial strain response of various quantum materials \cite{IkedaPNAS2021,LiNature2022,YePNAS2023,Gatinpj2023}. Uniaxial strain offers two primary advantages compared to doping-dependent studies to assess the evolution of multiple competing phases in a material. First, strain eliminates the need for multiple samples grown at slightly different dopings, which invariably host different levels of substitution-induced disorder \cite{DioguardiPRB2015}. Second, since the applied stress can be controlled precisely and also reversibly, it enables us to probe the phase diagram at a level of detail not achievable with chemical doping. Further, features beneath the superconducting dome which are invisible to resistivity measurements are accessible in ECE. Here, by comparing the ECE response between overdoped and slightly underdoped samples, we have uncovered thermodynamic evidence for a phase transition (at $T^*$) below \Tc on the underdoped side, which has remained undetected in previous studies. We find that our observations are most straightforwardly explained through the intertwining between the $s$ and $d$-wave SC instabilities, enabled by their close proximity and by the presence of nematic order. This leads to an $s+d$ order parameter condensing at \Tc, followed by time-reversal symmetry breaking at $T^*$ that is characterized by an $s+e^{i\phi}d$ OP, with the phase $\phi$ moving from $0, \pi$ towards $\pm\pi/2$ below $T^*$.

\section{Experiment}

The elastocaloric effect measures the temperature changes $\delta T$ induced by an applied strain $\delta\epsilon$ under adiabatic conditions. This is the strain analog of the traditional magnetocaloric effect, which measures the temperature changes induced by an applied magnetic field. The ECE directly probes the strain dependence of the entropy landscape ($\partial S/\partial\epsilon$) through the relation \cite{IkedaRSI2019}
\begin{equation}
	\frac{\delta T}{\delta \epsilon}\Big|_S=-\frac{T}{C_{\epsilon}}\cdot\frac{\partial S}{\partial \epsilon}\Big|_T,
\end{equation}
where $C_{\epsilon}$ is the specific heat under constant strain. Experimentally, we obtain the ECE response by applying a small AC stress to the sample and measuring the induced temperature oscillations $\delta T$ through standard lock-in techniques---see Ref. \cite{IkedaRSI2019} and SI for more experimental details. For all the experiments described here, uniaxial stress is applied along the crystalline [100] direction, that is, along the Fe-As bond direction. Note that this is different from the direction associated with nematic order, which is $[110]$, and thus a phase transition to a nematic state (relevant for underdoped compositions) is still allowed. The applied stress induces a longitudinal  strain $\epsilon_{xx}$, along with $\epsilon_{yy}=-\nu_{xy}\epsilon_{xx}$ and $\epsilon_{zz}=-\nu_{xz}\epsilon_{xx}$, where $\nu_{xy}$ and $\nu_{xz}$ are the in-plane and out-of-plane Poisson's ratios, respectively \cite{IkedaPRB2018}. The experiment is performed using a Razorbill CS100 strain cell \cite{HicksRSI2014}, which induces a given longitudinal strain based on the voltage applied to the driving piezoelectric stacks, and the strain $\epsilon_{xx}$ is measured by a capacitive sensor. In terms of irreducible strains \cite{irrep}, the sample experiences a combination of symmetric (\aog) and antisymmetric (\bog) strains, given by $\epsilon_{A_{1g},1}=\epsilon_{xx}+\epsilon_{yy}$, $\epsilon_{A_{1g},2}=\epsilon_{zz}$ and $\epsilon_{B_{1g}}=\epsilon_{xx}-\epsilon_{yy}$. The measured ECE response $\delta T$ is thus the total response of the material to this combination of strains. The AC stress can be superimposed on top of a fixed DC stress to measure the ECE response at different applied uniaxial stresses.

\begin{figure*}[h!]
	\centering
	\includegraphics[width=0.99\linewidth]{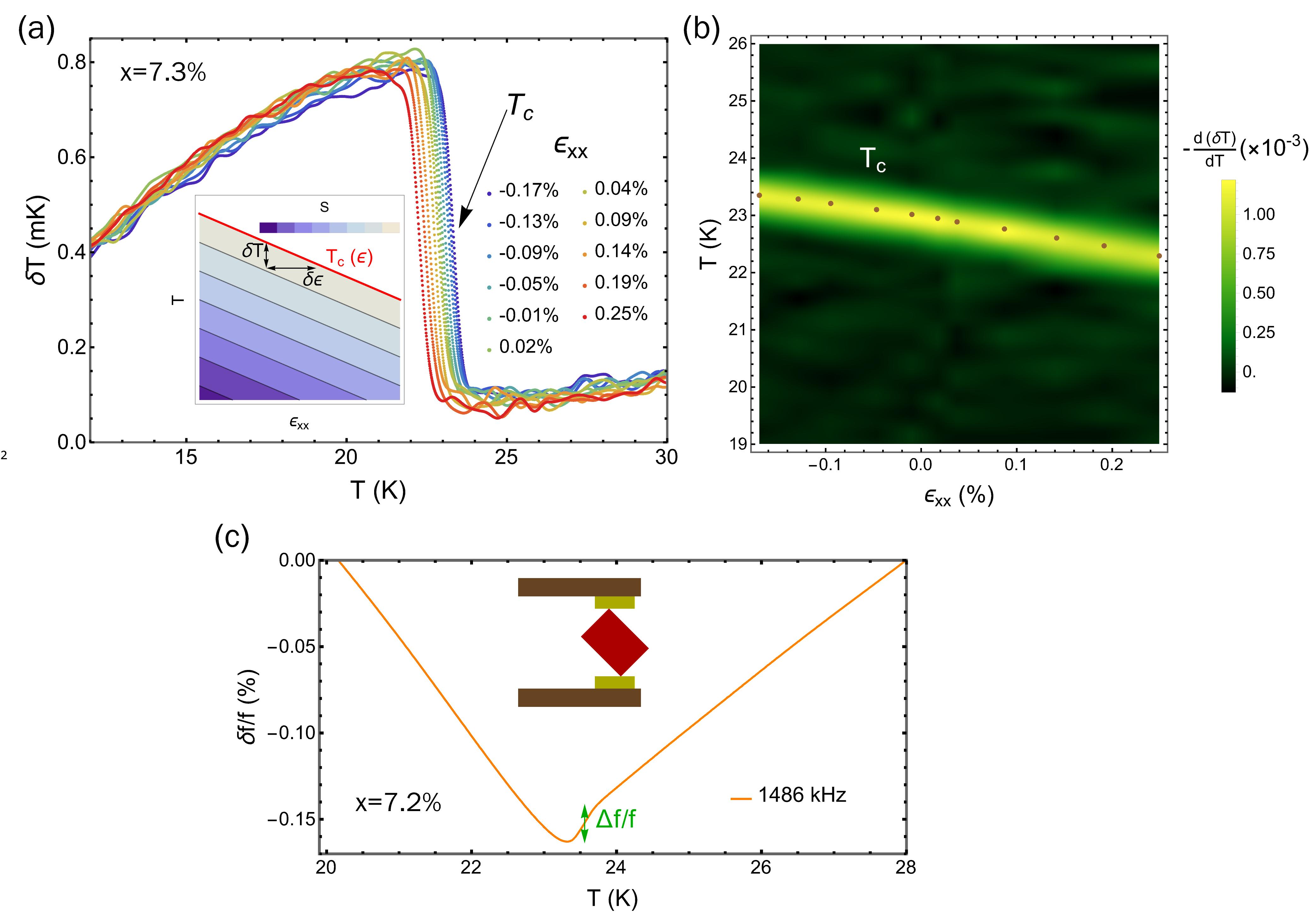}
	\caption{\label{fig:overdoped} \textbf{ECE and resonant ultrasound in overdoped Ba(Fe$_{1-x}$Co$_x$)$_2$As$_2$.} (a) ECE data for an overdoped ($x=7.3\%$) sample through \Tc for different bias strains $\epsilon_{xx}$ (strain along the Fe-As bond direction). Inset shows the expected entropy landscape in the temperature-strain plane proximate to a mean-field continuous transition when the critical temperature depends linearly on strain. Experimentally, the temperature changes $\delta T$ induced by an AC strain $\delta\epsilon$ are measured under adiabatic conditions ($\delta S=0$), as indicated by the arrows. (b) Temperature derivative of the data in (a) plotted on the $T-\epsilon_{xx}$ plane. The bright band highlights \Tc as a function of strain. Data points show the transition temperatures at experimentally measured strain values. (c) Temperature evolution of a primarily compressional ultrasonic resonance (related to the elastic modulus $\left(c_{11}+c_{12}\right)/2$) for an overdoped ($x=7.2\%$) sample through \Tc. The jump discontinuity at \Tc is highlighted. Inset shows a schematic of the ultrasound setup, see \citet{GhoshNatPhys2020} for more details.} 
\end{figure*}

\subsection{Overdoped}

We start with ECE measurements on the overdoped side of the phase diagram ($x=7.3\%$), where superconductivity condenses within the tetragonal phase in an $s$-wave state. The entropy landscape arising from the critical degrees of freedom near a superconducting transition, as shown in the inset of \autoref{fig:overdoped}(a), leads to a large $\partial S/\partial \epsilon$ at \Tc. This leads to a sharp, discontinuous feature in the ECE response ($\delta T$) as the sample becomes superconducting. As the external stress is varied from negative (compressive) to positive (tensile), the discontinuity moves to lower temperatures (\autoref{fig:overdoped} (a)). To extract the strain dependence of \Tc, we plot the derivative of $\delta T$ with respect to temperature $T$, which leads to the temperature-strain phase diagram plotted in \autoref{fig:overdoped}(b)---the bright band shows the superconducting phase boundary. We find that \Tc decreases monotonically with increasing $\epsilon_{xx}$, consistent with the fact that uniaxial strain $\epsilon_{xx}$ tunes the material in a similar fashion as Co doping \cite{MeingastPRL2012}.

Within a Landau theory, the linear dependence of \Tc on strain requires a coupling term of the form $|\Delta|^2\epsilon_\mu$, where $\Delta$ is the (complex) superconducting gap and $\epsilon_{\mu}$ is an irreducible strain. Any single-component superconducting gap can only couple to symmetric (\aog) strains as $|\Delta|^2\epsilon_{A_{1g}}$. Experimentally, we apply strain along the [100] direction ($\epsilon_{xx}$), which is a combination of symmetric (\aog) and antisymmetric (\bog) strains. Hence, this coupling is allowed and leads to the linear dependence of \Tc on $\epsilon_{xx}$. The linear dependence has a direct implication for the behavior of elastic moduli of the material through \Tc \cite{RamshawPNAS,GhoshSciAdv2020}. An allowed coupling of the form $|\Delta|^2\epsilon_{\mu}$ leads to a ``jump" discontinuity in the corresponding elastic modulus $c_{\mu}$ at \Tc. This elastic modulus discontinuity $\Delta c_{\mu}$ is related to the specific heat discontinuity $\left(\frac{\Delta C_p}{T_c}\right)$ and  $\frac{dT_c}{d\epsilon}$ through the Ehrenfest relation \cite{TestardiPRB1976} 
\begin{equation}
	\Delta c_{\mu}=-\frac{\Delta C_p}{T_c}\bigg(\frac{dT_c}{d\epsilon_{\mu}}\bigg)^2,
	\label{eq:ehrenModuli}
\end{equation}
where the negative sign denotes the opposite signs of the elastic modulus and specific heat discontinuities. Thus, a non-zero $dT_c/d\epsilon_\mu$ implies a jump in the corresponding elastic modulus $c_{\mu}$ at \Tc.

To confirm this, we performed resonant ultrasound spectroscopy (RUS) experiments on an overdoped sample, at a doping level close to the ECE sample. RUS measures the mechanical resonances of a solid, which are directly related to the elastic moduli of the solid---details of the experimental setup can be found in \citet{GhoshNatPhys2020}. The discontinuity in one of the resonances (1486 kHz) at \Tc is shown in \autoref{fig:overdoped}(c). After converting this frequency jump to an equivalent elastic moduli jump ($2\Delta f/f=\Delta c/c$, see Ref. \cite{GhoshNatPhys2020}), we can relate it to $dT_c/d\epsilon$ measured in ECE through \autoref{eq:ehrenModuli} \cite{RUSjump}. Our experimentally measured modulus jump $\Delta c=-(0.028\pm0.008)$ GPa compares well with the $\Delta c=-(0.0271\pm0.0004)$ GPa calculated from $\frac{dT_c}{d\epsilon}=-(244\pm2)$ K and $\frac{\Delta C_p}{T_c}=28$ mJ/mol/K$^2$ (reported in \citet{BudkoPRB2009} for a 7.4$\%$ doped sample). This shows quantitative consistency between different thermodynamic probes of the superconducting transition, namely, specific heat, elastic moduli and elastocaloric effect, on the overdoped side of the phase diagram.

\subsection{Slightly underdoped}

\begin{figure*}[h!]
	\centering
	\includegraphics[width=0.99\linewidth]{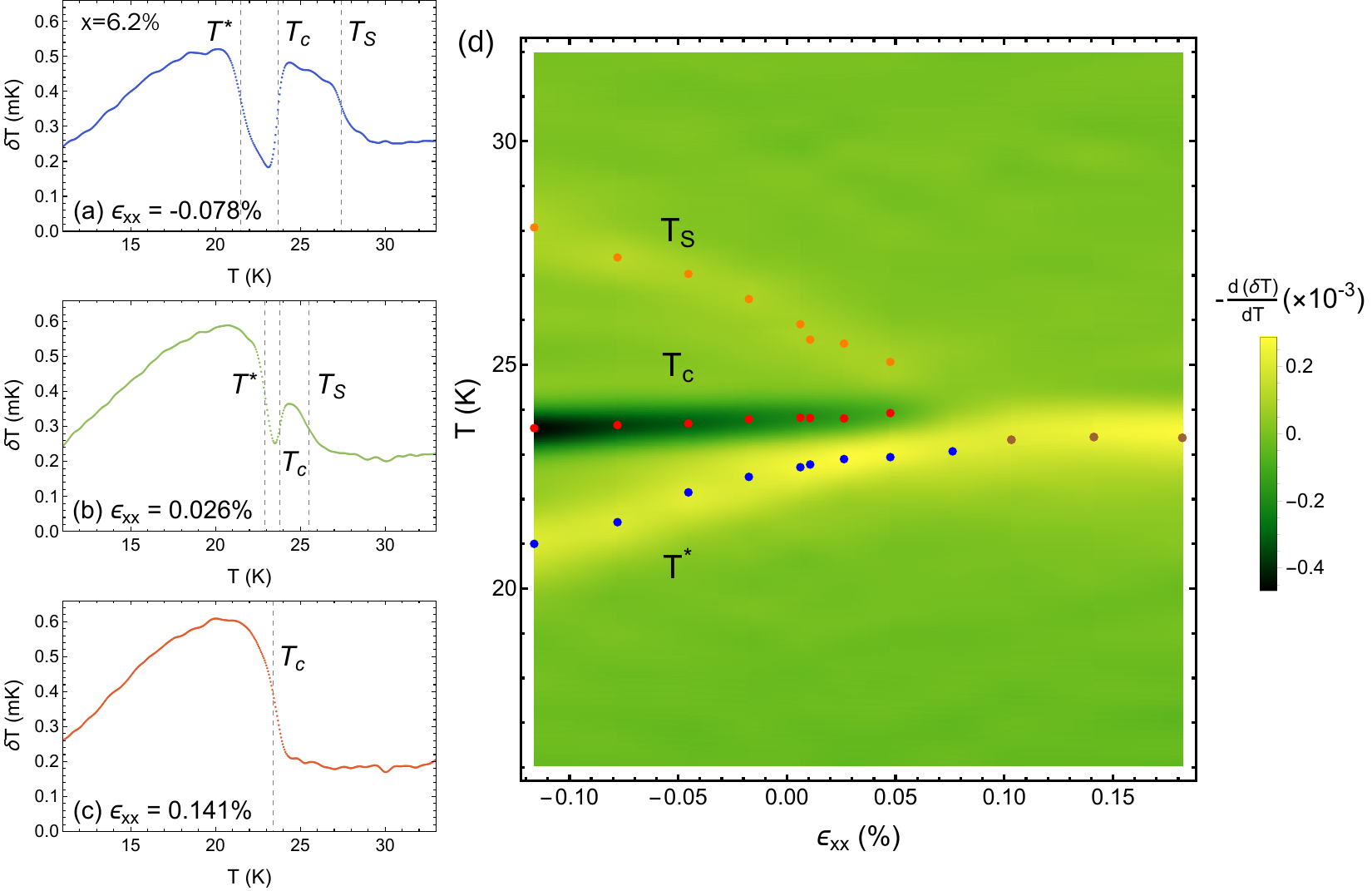}
	\caption{\label{fig:underdoped} \textbf{ECE in slightly underdoped Ba(Fe$_{1-x}$Co$_x$)$_2$As$_2$.} ECE data for a slightly underdoped (6.2$\%$) sample under (a) compressive, (b) near zero and (c) tensile strain. The observed phase transitions are indicated by dashed lines. Three distinct transitions are observed in (a) and (b), whereas only a single feature corresponding to the superconducting transition is seen in (c). (d) The resultant phase diagram showing the evolution of the three transitions, \TS, \Tc and $T^*$, in the temperature-strain plane. Data points show the \TS (orange points), \Tc (red points) and $T^*$ (blue points) at experimentally measured strain values. Brown points show \Tc for strains where only the superconducting transition occurs.} 
\end{figure*}

With a detailed understanding of the ECE measurements on the overdoped side, we can now investigate what happens on the underdoped side, where superconductivity condenses within the nematic/orthorhombic phase. We consider a composition that is only just underdoped, such that there is no antiferromagnetism, and the only phase transitions are for the  nematic and superconducting phases. The ECE data for such a slightly underdoped ($x=6.2\%$) sample is shown in \autoref{fig:underdoped}. The sample first undergoes the nematic transition (marked as \TS), where the tetragonal unit cell becomes orthorhombic. Since the unit cell distorts along the [110] direction, this remains a sharp transition even in the presence of the applied [100] stress. Around the superconducting transition, in stark contrast to the overdoped ECE data, multiple discontinuous features are seen in the ECE, marked as \Tc and $T^*$. This implies two adjacent thermodynamic phase transitions---this second transition below \Tc is our central finding.    

We now focus on how strain along [100], $\epsilon_{xx}$, tunes these transitions. Under the largest compressive strain (\autoref{fig:underdoped}(a)), three clear transitions can be identified, marked by $T_S$, \Tc and $T^*$. The onset of superconductivity at \Tc is also known through the observation of Meissner screening in simultaneous mutual inductance measurements on this sample (see SI for details). On tuning towards positive (tensile) strains, these transitions move closer together, such that only a single superconducting transition remains above a certain tensile strain ($\epsilon_{xx}\approx 0.07\%$). The ECE discontinuity at large tensile strains (\autoref{fig:underdoped}(c)) looks similar to the data from the overdoped sample (see \autoref{fig:overdoped}(a)), consistent with the fact that uniaxial tension along [100] tunes the material towards the overdoped side. The experimentally determined temperature-strain phase diagram (\autoref{fig:underdoped}(d)) looks similar to the well-known temperature-doping phase diagram across optimal doping, with the additional $T^*$ transition line that we have uncovered and which was not observed in previous doping-dependent studies. 

\begin{figure*}[h!]
	\centering
	\includegraphics[width=0.99\linewidth]{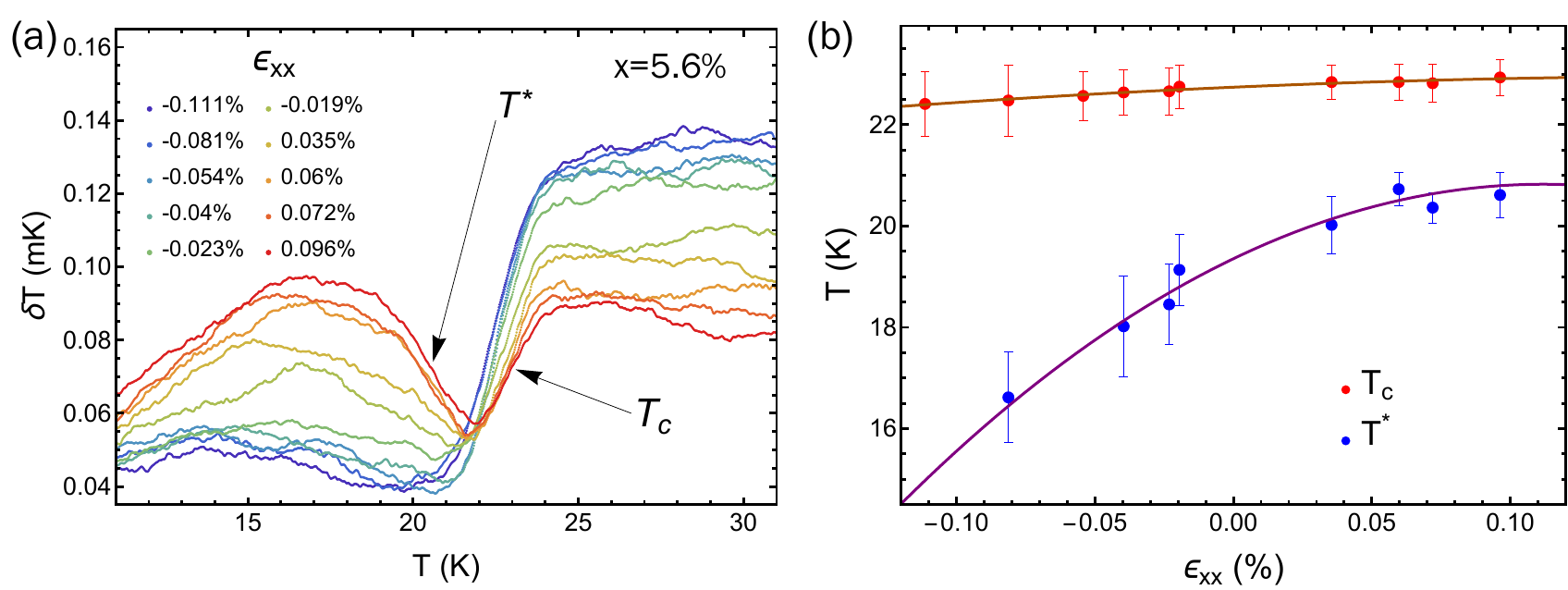}
	\caption{\label{fig:underdoped2} \textbf{Strain-dependence of \Tc and $T^*$.} (a) ECE data for a slightly underdoped (5.6$\%$) sample for different bias strains. (b) The extracted \Tc and $T^*$ as a function of strain are plotted. Error bars are derived from the width of the two transitions (see SI for details). Note the highly pronounced strain-dependence of $T^*$ compared to \Tc. The fits to $T_c(\epsilon)$ and $T^*(\epsilon)$ are obtained through relating the ECE discontinuities (from panel (a)) at these transitions to $dT_c/d\epsilon$ and $dT^*/d\epsilon$ through an Ehrenfest relation---see text and SI for details of the fitting procedure. The \Tc and $T^*$ lines remain separate, with the fits suggesting they would not cross at higher tensile strains.}
\end{figure*}

To further resolve the behavior of the \Tc and $T^*$ phase boundaries, we measured the ECE response from a second underdoped sample ($x=5.6 \%$) over a strain range where both \Tc and $T^*$ are observable (\autoref{fig:underdoped2}(a)). We find that the $T^*(\epsilon)$ line stays separate from $T_c(\epsilon)$ until the highest applied strains. The fits clearly suggest that they would not intersect (\autoref{fig:underdoped2}(b)), implying that the $T^*$ transition always occurs below \Tc, that is, within the superconducting state. The expected behavior of the three transitions near optimal doping, \TS, \Tc and $T^*$, as indicated by their evolution under uniaxial strain $\epsilon_{xx}$, is schematically shown in the temperature-doping phase diagram in \autoref{fig:phasediag}.

We further use the data in \autoref{fig:underdoped2}(a) to perform an important consistency check that relates the magnitude of ECE discontinuities to the strain derivative of the transition temperatures, $dT_c/d\epsilon$ and $dT^*/d\epsilon$. This arises due to the Ehrenfest relation (see Ref. \cite{YePNAS2023} for a derivation), 
\begin{equation}
	\Delta C_p\frac{dT_{\mathrm{crit}}}{d\epsilon}=\Delta\left[C_p\frac{\delta T}{\delta \epsilon}\right]\implies \frac{dT_{\mathrm{crit}}}{d\epsilon}=\frac{\Delta\left[C_p\frac{\delta T}{\delta \epsilon}\right]}{\Delta C_p},
	\label{eq:ehrenECE}
\end{equation} 
where $C_p$ denotes specific heat and $T_{\mathrm{crit}}$ is the transition temperature. The right-hand side of \autoref{eq:ehrenECE} can be calculated from the measured ECE data if the specific heat is known. Since this quantity gives $dT_{\mathrm{crit}}/d\epsilon$, $T_{\mathrm{crit}}(\epsilon)$ can be obtained by integrating the right-hand side. The fits to $T_c(\epsilon)$ and $T^*(\epsilon)$, shown as solid lines in \autoref{fig:underdoped2}(b), are obtained through this procedure---further details can be found in the SI. The goodness of the fits confirms the self-consistency of the data for both transitions. It also establishes that the second transition we have discovered at $T^*$ indeed behaves like a continuous thermodynamic transition.

From the above Ehrenfest relation, we estimate the expected specific heat discontinuity at the $T^*$ transition. This is particularly important since specific heat measurements only show a single discontinuity ($\Delta C_p$) at \Tc, with no signatures at $T^*$. Our measured specific heat on a similar doping sample also shows a single discontinuity (see SI), consistent with existing reports and ruling out any sample-specific issues. From the fit to $T^*\left(\epsilon\right)$ in \autoref{fig:underdoped2}(b), we extract a specific heat discontinuity at $T^*$, $\Delta C_p|_{T^*}$, which is approximately $3\% $ of the magnitude of $\Delta C_p|_{T_c}$. Intuitively, this follows from \autoref{eq:ehrenECE} and the fact that $dT^*/d\epsilon$ is about a factor of 11 larger than $dT_c/d\epsilon$, while the ECE discontinuities at \Tc and $T^*$ are comparable in magnitude (\autoref{fig:underdoped2}(a)). The smallness of $\Delta C_p|_{T^*}$ explains why, experimentally, a specific heat discontinuity is not resolved at $T^*$. In fact, the large value of $dT^*/d\epsilon$, which implies a strong sensitivity to strain, makes ECE the ideal probe to detect this transition, which has remained hidden to previous investigations.

In certain families of underdoped pnictides, an orthorhombic-to-tetragonal transition has been observed in the superconducting phase \cite{NandiPRL2010,WangPRB2016}. In electron-doped 122 compounds, this was attributed to the back-bending of  the nematic transition line due to the competition between nematic and superconducting phases. In hole-doped 122 compounds, this was attributed to the so-called $C_4$ magnetic phase \cite{Allred2016}. To check for re-entrant tetragonality, we performed X-ray diffraction (XRD) measurements on a slightly underdoped sample where zero-strain \TS, \Tc and $T^*$ are known from resistivity and ECE measurements (details of the XRD setup can be found in the SI). The X-ray data shows the onset of orthorhombicity at \TS, with the sample remaining orthorhombic down to the lowest measured temperature ($\sim12$ K), which is about 7 K lower than $T^*$. This rules out re-entrant tetragonality or a competing magnetic phase as the origin of the phase transition at $T^*$. Furthermore, in ultrasound experiments on an underdoped sample, the sound waves are completely attenuated below \TS due to the formation of orthorhombic domains \cite{BellPhysRev1963}. No sharp resonances are resolved even at $10$ K, which indicates that the sample remains orthorhombic down to this temperature.

\section{Discussion}

\begin{figure*}[h!]
	\centering
	\includegraphics[width=0.99\linewidth]{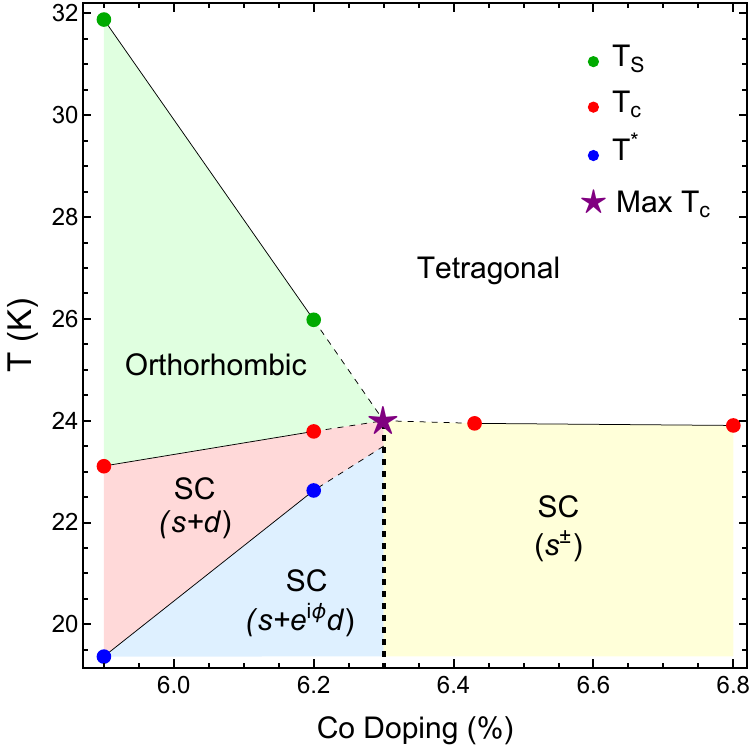}
	\caption{\label{fig:phasediag} \textbf{Phase diagram near optimal doping.} Temperature-doping phase diagram of Co-doped \bfs near optimal doping deduced from elastocaloric measurements. No antiferromagnetic order is observed in this doping range. Solid lines represent second order phase boundaries, with the dashed lines showing the expected behavior of \TS, \Tc and $T^*$ as optimal doping is approached. The thick dashed line represents a possibly first-order phase boundary that separates the superconducting states on the underdoped and overdoped sides. Note that the superconducting pairing symmetry changes across the first-order boundary.} 
\end{figure*}

By ruling out re-entrant tetragonality, we now discuss a possible scenario for the origin of the $T^*$ transition. Several theoretical microscopic models have predicted a $d_{xy}$-wave (in the 2-Fe crystallographic unit cell) superconducting instability to be nearly degenerate with the leading $s^\pm$-wave superconductivity near optimal doping (see Ref. \cite{HirschfeldIOP2011,ChubukovARCMP2012} for a comprehensive review). Such a near-degeneracy arises from the multi-orbital nature of the material combined with the different interactions that couple electronic states from distinct bands. Within the nematic phase, an $s+d$ order parameter must arise since the nematic order (\btg symmetry) couples the $s$ (\aog) and $d$ (\btg) superconducting OPs. Provided that these SC instabilities are close by in energy, the admixture between the $s$-wave and $d$-wave components in the nematic phase is not only large, but a second transition may also occur towards a time-reversal symmetry-breaking (TRSB) $s+e^{i\phi}d$ state, with $\phi\neq 0,\pi$ \cite{KangPRB2018}. Therefore, near but below optimal doping, where superconductivity condenses within the nematic state, we argue that an $s+d$ superconducting state first condenses at \Tc, followed by a TRSB transition at $T^*$ (see \autoref{fig:phasediag}). This scenario is consistent with the observation of a sub-leading $d$-wave Bardassis-Schrieffer collective mode in Raman spectroscopy \cite{MuschlerPRB2009,KretzschmarPRL2013} and the anisotropic gap deduced from thermal conductivity measurements \cite{TanatarPRL2010}.

To formalize this scenario, we consider a phenomenological mean-field Landau theory of competing $s$ and $d$-wave superconducting OPs , $\Delta_s$ and $\Delta_d$, with nearby transition temperatures (see SI for details). In the absence of nematic order, if both OPs condense, their relative phase $\phi$  is determined by the biquadratic term $u |\Delta_s|^2 |\Delta_d|^2 \cos 2\phi$, which generally favors a fully gapped $s+id$ state, i.e. $\phi=\pm\pi/2$ ($u>0$) \cite{KhodasPRL2012,PlattPRB2012}.  The presence of nematic order, however, generates a new bilinear coupling $g\Delta_s\Delta_d\cos\phi$, which favors instead a relative phase $\phi=0,\pi$  \cite{FernandesPRL2013}. Due to the existence of these two competing terms, this model can display two sequential superconducting transitions, as discussed in Ref. \cite{KangPRB2018}. At \Tc, the bilinear term wins and superconductivity condenses with $\Delta_{s,d}\neq0$ and $\phi=0, \pi$, which is the time-reversal invariant $s+d$ state. Note that, because both $s$ and $d$ OPs transform in the same way under the symmetry operations of the orthorhombic phase, this SC state does not break any additional symmetries. As the temperature is lowered and the gaps increase, the biquadratic term becomes more important relative to the bilinear term. As a result, for a range of parameters of the model, the relative phase $\phi$ unlocks from its \Tc value at  $T^*<T_c$, leading to a time-reversal symmetry breaking (TRSB) transition into an $s+e^{i\phi}d$ state (with $\phi\neq0,\pi$) \cite{KangPRB2018}.

To directly relate this theory to our experiments, we calculate the ECE discontinuities predicted by this model (details in the SI) and show that the ECE signatures at both \Tc and $T^*$ can be qualitatively reproduced. This is particularly non-trivial since $dT_c/d\epsilon$ and $dT^*/d\epsilon$ are both positive (\autoref{fig:underdoped2}), but the signs of the ECE discontinuities, which, for completely independent order parameters, are expected to follow the signs of $dT_{\mathrm{crit}}/d\epsilon$, are opposite at \Tc and $T^*$. We find that, because both the $s$ and $d$-wave OPs are already non-zero at $T^*$, a TRSB transition from the $s+d$ to the $s+e^{i\phi}d$ state naturally allows for the ECE discontinuity at $T^*$ to not follow the sign of $dT^*/d\epsilon$. Importantly, this is not allowed for a transition from a single-component ($s$ or $d$) to a multi-component SC state ($s+d$) at $T^*$. The opposite signs of the ECE discontinuities therefore strongly indicates that the lower transition is due to TRSB. This conclusion motivates further work to directly probe the existence of broken time reversal symmetry. Muon spin rotation ($\mu$SR) measurements under uniaxial strain may provide direct experimental evidence of time-reversal symmetry breaking at $T^*$ \cite{GrinenkoNatPhys2021}. It should also be possible to train the $s+e^{i\phi}d$ state using a combination of uniaxial strain $\epsilon_{xx}$ and out-of-plane magnetic field to observe a non-zero polar Kerr signal. Preliminary Kerr effect measurements under a combination of uniaxial strain and magnetic field ($\sim30$ mT) do not show clear evidence for TRSB at $T^*$, though the absence of a signature does not preclude the existence of such a state and further experiments are ongoing.

In summary, our work establishes the presence of a symmetry-breaking transition within the superconducting state in slightly underdoped Ba(Fe$_{1-x}$Co$_x$)$_2$As$_2$. The observation of this highly strain-sensitive transition was facilitated by using the AC elastocaloric effect, which measures the strain derivative of the entropy ($\partial S/\partial\epsilon$) and consequently detects a large signal at this transition. The presence of this transition modifies the zero-strain doping phase diagram (shown in \autoref{fig:phasediag}) and emphasizes the need for re-examining the superconducting pairing state in this extensively studied material. It also indicates two crucial roles of the nematic order in affecting superconductivity. First, the nematic phase appears to be a key determinant for the SC pairing symmetry, rendering the SC order parameter different in the orthorhombic phase compared to the tetragonal phase. This suggests a possible scenario where the nematic transition ends in a line of first order transitions below the SC dome (see \autoref{fig:phasediag}). Second, \Tc appears to be maximized when the nematic transition line meets the superconducting dome, which defines the optimal doping point in the phase diagram. Given the prevalence of a nematic phase among several families of Fe-based superconductors \cite{ShibauchiARCMP2014,KuoScience2016,IshidaPNAS2022}, whether the existence of a sub-dominant pairing channel is ubiquitous among them remains an intriguing question for future studies.

\section{Acknowledgements}

This work was supported by the Department of Energy, Office of Basic Energy Sciences, under contract DE-AC02-76SF00515. S.G. was partially supported by the Gordon and Betty Moore Foundation EPiQS Initiative, grant GBMF9068. Part of this work was performed at the Stanford Nano Shared Facilities (SNSF), supported by the National Science Foundation under award ECCS-2026822. A.R.C. and R.M.F. (theoretical model) were supported by the U. S. Department of Energy, Office of Science, Basic Energy Sciences, Materials Sciences and Engineering Division, under Award No. DE-SC0020045. This research used resources of the Advanced Photon Source, a U.S. Department of Energy (DOE) Office of Science user facility and is based on work supported by Laboratory Directed Research and Development (LDRD) funding from Argonne National Laboratory, provided by the Director, Office of Science, of the U.S. DOE under Contract No. DE-AC02-06CH11357.


\newpage

\section*{Supplementary Information}

\subsection*{Sample Characterization}

\begin{figure*}[h!]
	\centering
	\includegraphics[width=0.95\linewidth]{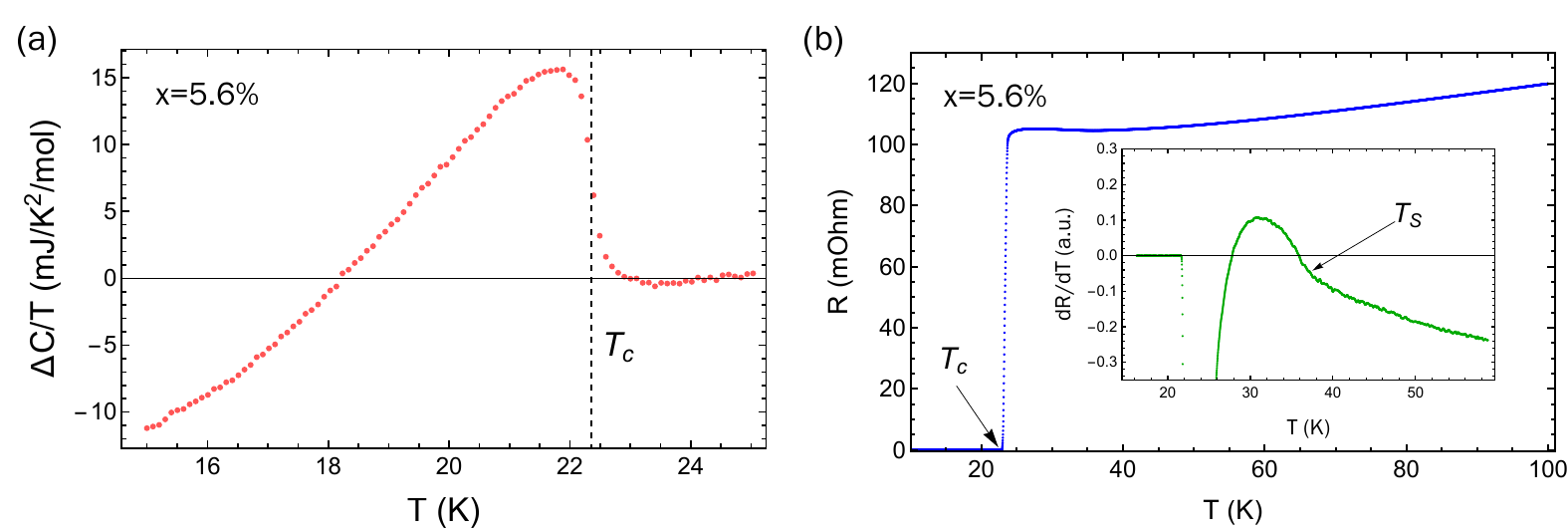}
	\caption{\label{fig:char} \textbf{Characterization measurements on slightly underdoped ($x=5.6\%$) samples.} (a) Specific heat of the sample through \Tc. A clear discontinuity is seen at \Tc, but no other features are observed. (b) Resistance measured as a function of temperature shows the structural (\TS) and superconducting (\Tc) transitions. Inset shows the temperature derivative of resistance, from which we extract \TS.}
\end{figure*}

Co-doped \bfs single crystals were grown by a self-flux technique described in \citet{ChuPRB2009}. Heat capacity and resistance measurements were performed to find the structural and superconducting transition temperatures. We show the corresponding data for two slightly underdoped samples ($x=5.6\%$) in \autoref{fig:char}. Heat capacity (\autoref{fig:char}(a)) shows a clear discontinuity at \Tc$\approx22.4$ K, and the size of the discontinuity $\Delta C_p$ compares well with that reported in Ref. \cite{ChuPRB2009}. Resistance, measured by the standard 4-point method, clearly shows the expected feature at \TS (shown in inset of \autoref{fig:char}(b)) and exhibits \Tc$\approx22.5$ K, when resistance goes to zero. 

\begin{figure*}[h!]
	\centering
	\includegraphics[width=0.95\linewidth]{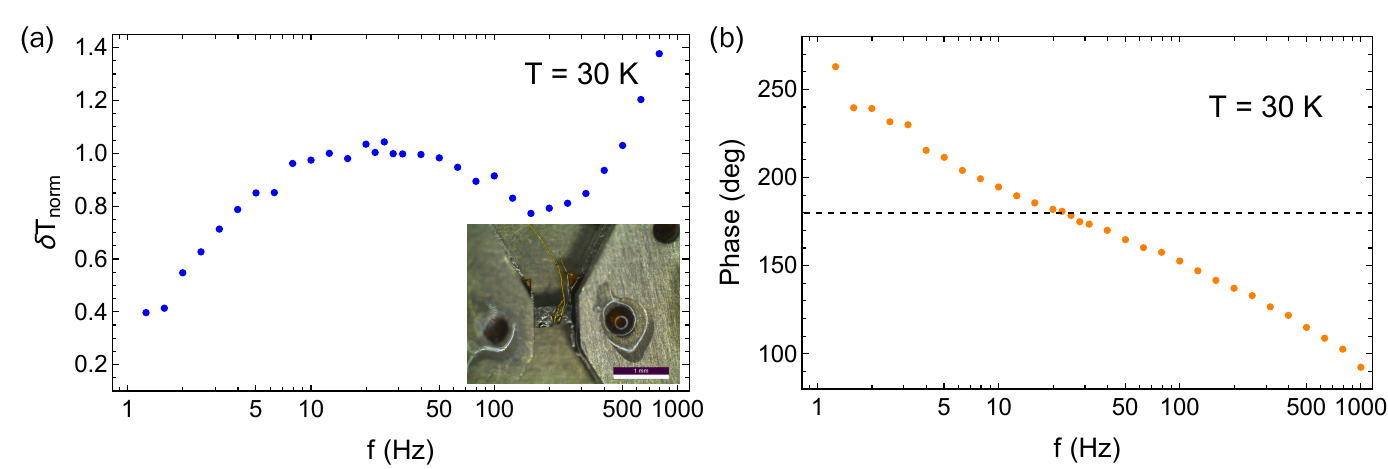}
	\caption{\label{fig:fscan} \textbf{ECE frequency scan.} (a) The amplitude of elastocaloric temperature oscillations measured by the thermocouple as the frequency $f$ is varied. The data is normalized to $\delta T$ at $\sim 13$ Hz, which is the frequency at which the experiment is performed. Inset shows a photograph of the sample held within the strain cell. (b) The phase of ECE response as the frequency is varied. The dashed line marks where the phase crosses 180 degrees.}
\end{figure*}

\subsection{ECE experimental details and additional data}

ECE measurements at cryogenic temperatures were performed using a Razorbill CS-100 strain cell. The sample is attached to Ti mounting plates with DEVCON 2-ton Epoxy and cured at room temperature. The samples used for ECE measurements are shaped into long thin bars with typical dimensions 1.80$\times$0.35$\times$0.03 mm, with strain being applied along the long direction. Care is taken to have at least 300 $\mu$m of the sample held by the plates on either side, this ensures good strain transmission into the sample and reduces chances of the sample slipping during cooldown. To measure the ECE temperature oscillations, we use a chromel-Au (0.07$\%$ Fe) thermocouple, which is attached to the sample using a thin layer of AngstromBond AB9110LV. A sample mounted in the strain cell is shown in the inset of \autoref{fig:fscan}(a).

It is important to determine the experimental frequency such that most of the AC strain-induced temperature changes are sensed by the thermocouple. This has to be determined separately for each different sample, since the optimal frequency may depend on sample dimensions, amount of glue, etc. This frequency is fixed by the competition between two timescales \cite{IkedaRSI2019}: the time taken for heat to flow out of the sample into the Ti plates ($\tau_b$) and the time taken for the thermocouple to thermalize with the sample ($\tau_\theta$). At low enough experimental frequencies ($\tau_b^{-1}\gg f$), most of the heat flows out of the sample and thus the temperature changes induced at the thermocouple are small. On the other hand, at high frequencies ($\tau_\theta^{-1}\ll f$), the thermocouple cannot thermalize with the sample and thus the response is small. We thus need to experimentally determine an appropriate ``quasi-adiabatic" frequency ($\tau_b^{-1}<f<\tau_\theta^{-1}$), such that most of the intrinsic elastocaloric temperature change is sensed by the thermocouple. To find this, we vary the frequency of applied stress and record the thermocouple response at different frequencies  (see \autoref{fig:fscan}). A clear plateau in the ECE response $\delta T$ is observed around 20 Hz, indicating an optimal frequency in this regime. The corresponding phase is also seen to cross $\pi$ or 180 degrees at the same frequency, which points to the thermocouple sensing temperature changes in-phase with the applied stress. The fact that the phase is $\pi$ and not zero arises due to the sign of ECE in this sample, that is, positive strain (tension) leads to cooling of the sample (negative $\delta T$). The rise in $\delta T_{norm}$ above $\sim200$ Hz is extrinsic and possibly related to mechanical vibrations of the piezoelectric stacks within the strain cell \cite{YePNAS2023} or heating of the piezoelectric stack due to rapidly varying voltages. We performed the experiments for this sample at an AC strain frequency of 13.333 Hz. Typical AC strain oscillation amplitude used in the experiments is approximately 5$\times10^{-5}$. 

\begin{figure*}[h!]
	\centering
	\includegraphics[width=0.75\linewidth]{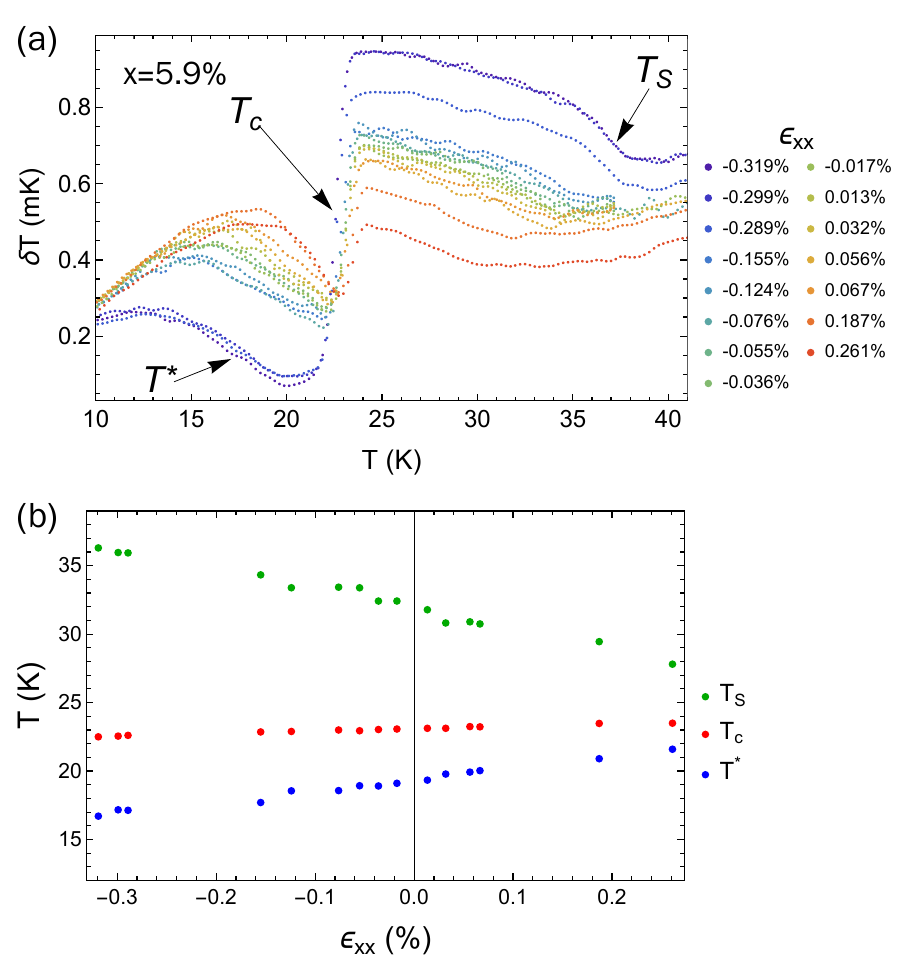}
	\caption{\label{fig:ECE5p9} \textbf{ECE data on a $x=5.9\%$ sample.} (a) ECE data at different applied strains. (b) Evolution of the transition temperatures \TS, \Tc and $T^*$ with strain at this doping.}
\end{figure*}

\begin{figure*}[h!]
	\centering
	\includegraphics[width=0.99\linewidth]{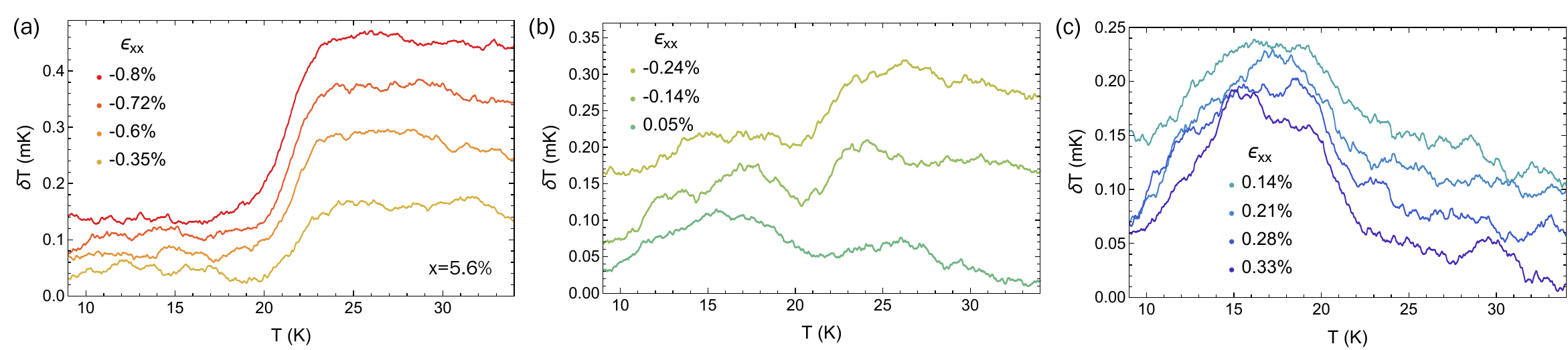}
	\caption{\label{fig:bigStr} \textbf{ECE data on a $x=5.6\%$ sample for large applied strains.} ECE data at (a) compressive, (b) near zero, and (c) tensile stress on a single sample. Relatively large compressive and tensile strains were applied to this sample. Plots are shifted vertically for visual clarity. The sign of the ECE discontinuity at \Tc changes as the sample goes from underdoped-like (panel (a)) to overdoped-like (panel (c)). At intermediate strains, two discontinuities, corresponding to \Tc and $T^*$, are observable in the data.}
\end{figure*}

ECE data in two additional slightly underdoped compositions are shown in \autoref{fig:ECE5p9} and \autoref{fig:bigStr}. Three clear transitions, corresponding to \TS, \Tc and $T^*$, are seen in \autoref{fig:ECE5p9}(a). The evolution of the three transition temperatures as a function of applied strain compares well with the behavior of the $x=6.2\%$ sample shown in Main Text Figure 2. We use the zero strain values of \TS, \Tc and $T^*$ from \autoref{fig:ECE5p9}(b) to construct the phase diagram shown in Main Text Figure 4. To study the behavior of \Tc and $T^*$ over a large range of strains, we measured the ECE response from a $x=5.6\%$ sample, shown in \autoref{fig:bigStr}. To apply large strains in this sample, we used a half-cut Ti plate on one side for mounting, which allowed us to strain $\sim 500$ $\mu$m of the sample, compared to $\sim800$ $\mu$m for usual samples. The ECE discontinuity at \Tc changes sign as we tune from large compressive strains (\autoref{fig:bigStr}(a)) to large tensile strains (\autoref{fig:bigStr}(c)). This is consistent with the sample going from underdoped-like to overdoped-like under [100] uniaxial tension. For intermediate strains (\autoref{fig:bigStr}(b)), two clear transitions, at \Tc and $T^*$, are observed. The overall behavior is consistent with, and further confirms, the temperature-strain phase diagram we plot in Main Text Figure 2.

\begin{figure*}[h!]
	\centering
	\includegraphics[width=0.6\linewidth]{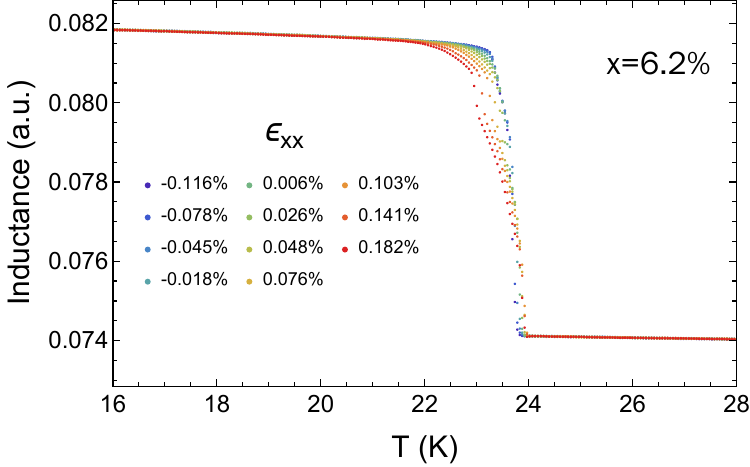}
	\caption{\label{fig:induct} \textbf{Mutual inductance showing onset of \Tc.} Mutual inductance measured on a $x=6.2\%$ sample simultaneously with ECE. The ECE data is shown in Main Text Figure 2. From the jump in inductance, we confirm that the higher temperature feature in ECE is due to the superconducting transition.}
\end{figure*}



\subsection{X-ray diffraction measurements}

\begin{figure*}[h!]
	\centering
	\includegraphics[width=0.6\linewidth]{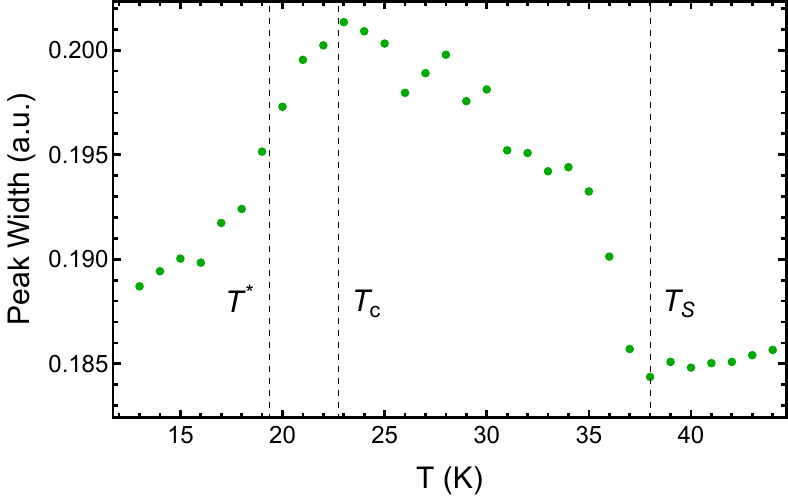}
	\caption{\label{fig:xray} \textbf{X-ray diffraction on a slightly underdoped ($x=5.6\%$) sample.} Width of the (224) Bragg peak measured with X-ray diffraction. Clear increase in the peak width is seen at \TS and the width starts decreasing at \Tc. The peak width below $T^*$ shows that the sample stays orthorhombic below $T^*$. The \TS, \Tc and $T^*$ are determined from separate resistivity and ECE measurements.} 
\end{figure*}

To check for orthorhombicity, we performed X-ray diffraction (XRD) measurements on a slightly underdoped ($x=5.6\%$) sample where zero-strain \TS, \Tc and $T^*$ are known from resistivity and ECE measurements. The XRD data was measured using a Huber Psi-circle diffractometer equipped with a 4K ARS closed-cycle He cryostat and using a Pilatus 100K area detector. The incoming X-rays were produced by a Xenocs GeniX 3D source at the Cu K$_\alpha$ line. At each temperature, a rocking curve measurement of the (224) Bragg peak was measured. The resulting detector images were binned into the scattering angle (2$\theta$), leading to intensity profiles in which the main signal peak was fit by a Gaussian peak. The full width at half maximum (FWHM) as a function of temperature is shown in \autoref{fig:xray}. We see a clear increase in the FWHM around \TS$=38$ K, indicating the tetragonal-to-orthorhombic transition. The increase in the width mimics the growth of the nematic order parameter and it starts decreasing around \Tc$=22$ K, when superconductivity condenses and competes with nematicity. However, no sharp feature is observed at $T^*\approx19$ K, and the sample stays orthorhombic till the lowest temperatures ($\sim12$ K). Since the peak width is a direct evidence for orthorhombicity, its temperature dependence and the fact that it behaves as expected at \TS and \Tc allows us to rule out re-entrant tetragonality as the origin of the phase transition at $T^*$.

\begin{figure*}[h!]
	\centering
	\includegraphics[width=0.6\linewidth]{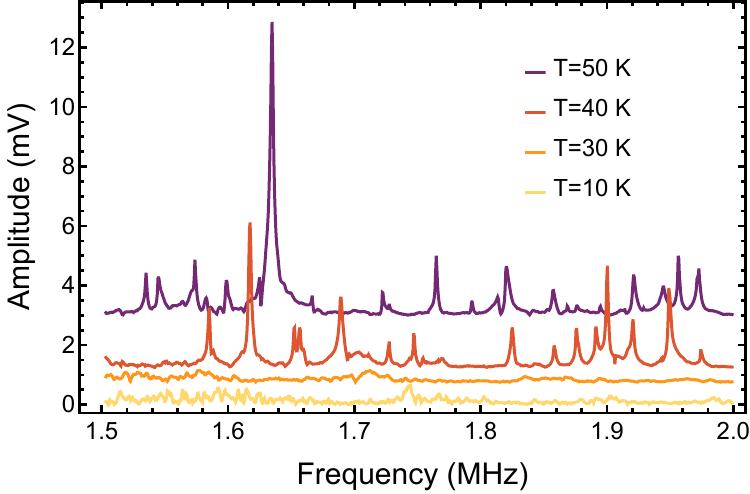}
	\caption{\label{fig:RUS} \textbf{Ultrasound data on a slightly underdoped ($x=5.6\%$) sample.} RUS frequency sweeps showing the resonance frequencies of the sample between 1.5-2.0 MHz. The sharp resonances, seen at 40 K and 50 K, disappear in the orthorhombic phase due to strong attenuation by domains. The resonances do not reappear even at 10 K.} 
\end{figure*}

\subsection*{Resonant ultrasound spectroscopy (RUS) measurements}

RUS measures the mechanical resonances of a three-dimensional solid---these resonance frequencies are related to the various independent elastic moduli of the solid. The sample is placed in weak mechanical contact with two piezoelectric transducers. An AC excitation is applied to one of the transducers to generate a ultrasonic wave within the sample. The voltage sensed on the second transducer at the same frequency shows a peak when the excitation frequency matches a mechanical resonance of the sample. We use a custom-made low temperature RUS setup for the experiments reported here \cite{GhoshNatPhys2020}. The samples used were single crystals cut into thin rectangular plate-like shapes. The dimensions of the overdoped sample ($x=7.3\%$) is 1.40$\times$1.14$\times$0.19 mm, with 0.19 mm along the tetragonal c-axis. The dimensions of the underdoped sample ($x=5.6\%$) is 1.69$\times$1.51$\times$0.23 mm, with 0.23 mm along the tetragonal c-axis. 

RUS frequency sweeps for the underdoped sample at a few representative temperatures are shown in \autoref{fig:RUS}. Clear, sharp resonances are seen at 50 K and 40 K, which abruptly disappear once the sample becomes orthorhombic below \TS$=38$ K. This is due to the formation of structural domains which scatter ultrasound---similar behavior is observed across structural transitions in other materials \cite{BellPhysRev1963}. No sharp resonances are resolved even at 10 K, which indicates the domains persist down to this temperature. This provides additional evidence for the sample remaining orthorhombic at this temperature and rules against a re-entrant tetragonal phase at $T^*\approx19$ K.

\subsection*{Defining transition temperatures from ECE data}

\begin{figure*}[h!]
	\centering
	\includegraphics[width=0.95\linewidth]{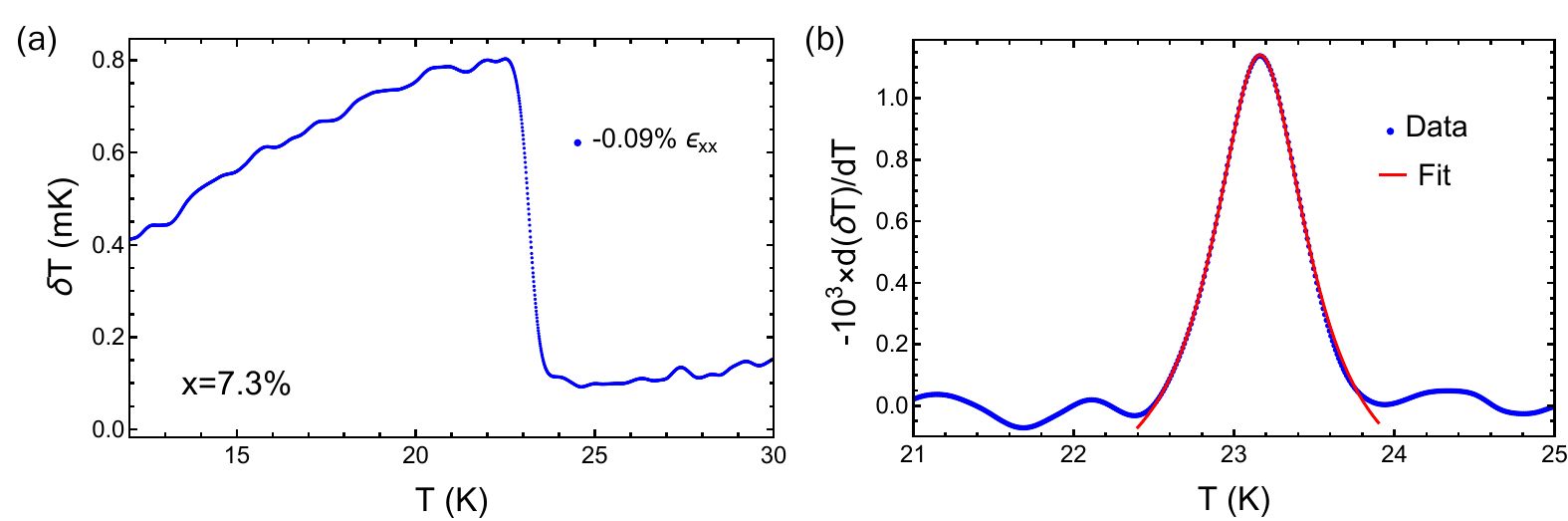}
	\caption{\label{fig:ECEder1} \textbf{Extracting transition temperature from a temperature derivative.} (a) ECE data for the overdoped sample at a particular strain ($\epsilon_{xx}=-0.09\%$). (b) Temperature derivative of the data in (a), averaged over a 100 mK window. A clear peak shows up indicating \Tc, which we fit to a Lorentzian function to extract \Tc at this strain value.}
\end{figure*}

A sharp discontinuous feature in the elastocaloric measurements indicates a phase transition (\autoref{fig:ECEder1}(a)). To extract the transition temperature(s), we take a temperature derivative of the elastocaloric data. The sharp discontinuity then shows up as a peak in the derivative, which allows defining the transition temperature very reliably. We fit a Lorentzian to this peak, $a_0+\frac{A}{(T-T_0)^2+B^2}$, from which we obtain $T_0$ as the transition temperature (\autoref{fig:ECEder1}(b)). The error bar on the transition temperature, in this case, comes from the fit error on $T_0$. In certain datasets, when the size of the discontinuity is small or the transition is somewhat broadened, this procedure does not work well. In those cases, we define the transition temperature as the midpoint of the discontinuous feature, which can be identified by visual inspection alone. The error bars on the transition temperatures are then defined as 30$\%$ of the width of the discontinuity. This gives the error bars on \Tc and $T^*$ shown in Main Text Figure 3.

\subsection*{Strain-Order Parameter Coupling in an $s$-wave superconductor}

Overdoped \bfs has a single component, $s_\pm$, superconducting (SC) gap. For a $s$-wave SC order parameter, the free energy near the SC transition can be written as,
\begin{equation}
	\mathcal{F}_{op}(\Delta_s) =a|\Delta_s|^2+b|\Delta_s|^4
	\label{eqn:Fop}
\end{equation}
where $a=a_0(T-T_c)$, with $a_0>0$, and $b>0$ is a constant.
The elastic free energy density is given by
\begin{equation}
	\mathcal{F}_{el}=\frac{1}{2}\sum_{\mu}c_{\mu}\epsilon_{\mu}^2,
\end{equation}
where $\epsilon_{\mu}$ are the strains in irreducible representations (irreps) \cite{GhoshNatPhys2020} and $c_\mu$ are the corresponding elastic moduli. The various strains in a tetragonal crystal denoted in terms of their irreps are: $(\epsilon_{xx}+\epsilon_{yy}) \rightarrow \epsilon_{A_{1g,1}}$, $\epsilon_{zz} \rightarrow \epsilon_{A_{1g,2}}$, $(\epsilon_{xx}-\epsilon_{yy}) \rightarrow \epsilon_{B_{1g}}$, $2\epsilon_{xy} \rightarrow \epsilon_{B_{2g}}$ and $(2\epsilon_{xz},2\epsilon_{yz}) \rightarrow \epsilon_{E_g}$.

The coupling between strains and $\Delta_s$ gives rise to additional contributions to the free energy. Note that the lowest order coupling term must be quadratic in SC order parameter to preserve gauge symmetry, leading to
\begin{equation}
	\mathcal{F}_{c} =(g_1\epsilon_{A_{1g,1}}+g_2\epsilon_{A_{1g,2}})|\Delta_s|^2,
	\label{eqn:Fc}
\end{equation}
where $g_i$ are coupling constants. These are the only terms allowed by symmetry for a $s$-wave superconducting gap.

The total free energy thus becomes,
\begin{equation}
	\mathcal{F}_{tot}=a_0(T-T_c)|\Delta_s|^2+b|\Delta_s|^4+\frac{1}{2}\sum_{\mu}c_{\mu}\epsilon_{\mu}^2+(g_1\epsilon_{A_{1g,1}}+g_2\epsilon_{A_{1g,2}})|\Delta_s|^2.
\end{equation}
The transition temperature is determined by setting the coefficient of $|\Delta_s|^2$ to zero, which gives
\begin{equation}
	T_c(\epsilon_{A_{1g,1}},\epsilon_{A_{1g,2}})=T_c-\frac{g_1}{a_0}\epsilon_{A_{1g,1}}-\frac{g_2}{a_0}\epsilon_{A_{1g,2}}.
\end{equation}
This shows that coupling terms of the form $\Delta_s^2\epsilon$ leads to a linear variation of \Tc with strain. Such a coupling further leads to a discontinuity in the corresponding modulus $c_\mu$ \cite{SigristPTP2002}, given by
\begin{equation}
	\delta c_{A_{1g,1(2)}}=-\frac{g_{1(2)}^2}{b}.
\end{equation} 
The specific heat discontinuity (at zero strain) at \Tc is,
\begin{equation}
	\frac{\delta C_p}{T_c}=\frac{a_0^2}{b}.
\end{equation}
Clearly, the above three equations satisfy the Ehrenfest relation
\begin{equation}
	\delta c_{A_{1g,1(2)}}=-\frac{\delta C_p}{T_c}\bigg(\frac{dT_c}{d\epsilon_{A_{1g,1(2)}}}\bigg)^2,
\end{equation}
which we verify for the overdoped sample from elastocaloric, specific heat and ultrasound measurements.

\subsection*{Theoretical calculation of the elastocaloric effect in an $s+\mathrm{e}^{i\phi} d$ superconductor}
The ECE coefficient $\eta$, whose value denotes the adiabatic change in temperature as function of strain, is defined as (see e.g., Ref. \cite{YePNAS2023}) 
\begin{equation}
	\eta = \Big(\frac{\partial T}{\partial \epsilon} \Big)_S \label{eq:def_eta}
\end{equation}
In our case, the entropy $S$ is a function of temperature $T$ and strain $\epsilon$,  which for now we may consider to be either symmetry-preserving or symmetry-lowering (shear) strain. It follows that the variation of $S$ is 
\begin{equation}
	dS=\Big(\frac{\partial S}{\partial T}\Big)_\epsilon dT + \Big(\frac{\partial S}{\partial \epsilon}\Big)_T d\epsilon
\end{equation}
Differentiation with respect to $T$ at constant $S$ yields a Maxwell relation, from which we write $\eta$ in terms of the derivatives of entropy:
\begin{equation}
	\eta = -\frac{(\partial S/\partial \epsilon)_T}{(\partial S/\partial T)_\epsilon}
\end{equation}
As shown in Ref. \cite{YePNAS2023}, the ECE coefficient satisfies an $\epsilon$-$T$ generalized Ehrenfest relation, which will be useful for our analysis. For completeness, we show the derivation of this result here. Let the temperature $T_0$ mark a second-order transition between a `$-$' phase for $T<T_0$ and a `$+$' phase for $T>T_0$. Owing to the continuity of the entropy across a second-order transition, an infinitesimal variation of the entropies of the two phases at the critical temperature $T_0$ must be equal: $dS_- = dS_+$. For $S=S(T,\epsilon)$ this condition becomes
\begin{equation}
	\Big(\frac{\partial S_-}{\partial T}\Big)_\epsilon dT + \Big(\frac{\partial S_-}{\partial \epsilon} \Big)_T d\epsilon=\Big(\frac{\partial S_+}{\partial T}\Big)_\epsilon dT + \Big(\frac{\partial S_+}{\partial \epsilon} \Big)_T d\epsilon
\end{equation}
from which it follows that
\begin{equation}
	\frac{dT_0}{d\epsilon} = -\frac{\delta(\partial S/\partial \epsilon)_{T_0}}{\delta(\partial S/\partial T)_{T_0}}
	\label{eq:dT0deps}
\end{equation}
after differentiating with respect to $\epsilon$ and setting $T=T_0(\epsilon)$. We define the discontinuous jump at $T_0$ as
$\delta(\partial S/\partial x)_{T_0}\equiv(\partial S/\partial x)_{T\to T_0^-} - (\partial S/\partial x)_{T\to T_0^+}$, with $x=T,\epsilon$. The notation $T\to T_0^\pm$ indicates that $T$ approaches $T_0$ from above or below, respectively. Note that the derivatives are taken with respect to the other thermodynamic variable held fixed.

We can now express $dT_0/d\epsilon$ in terms of $\eta$ and the constant-strain specific heat $C_\epsilon=T(\partial S/\partial T)_\epsilon$. Using Eq. (\ref{eq:def_eta}) to eliminate $(\partial S/\partial \epsilon)_{T_0}$  in Eq. (\ref{eq:dT0deps}) and rearranging the terms leads to the symmetric form of the Ehrenfest relation 
\begin{equation}
	\frac{dT_0}{d\epsilon} = \Bigg[\frac{\delta (\eta C_\epsilon)}{\delta C_\epsilon}\Bigg]_{T_0}=\frac{\eta(T_0^-)C_\epsilon(T_0^-)-\eta(T_0^+)C_\epsilon(T_0^+)}{C_\epsilon(T_0^-)-C_\epsilon(T_0^+)}.
	\label{eq:ehrenECE}
\end{equation}
Isolating the discontinuity of the ECE coefficient $[\delta\eta]_{T_0}=\eta(T_0^-)-\eta(T_0^+)$ gives:
\begin{equation}
	[\delta \eta]_{T_0} = \frac{[\delta C_\epsilon]_{T_0}}{C_\epsilon(T_0^+)}\Bigg[\frac{dT_0}{d\epsilon} - \eta(T_0^-) \Bigg]
	\label{eq:deltaECE}
\end{equation}
or, equivalently, 
\begin{equation}
	[\delta \eta]_{T_0} = \frac{[\delta C_\epsilon]_{T_0}}{C_\epsilon(T_0^-)}\Bigg[\frac{dT_0}{d\epsilon} - \eta(T_0^+) \Bigg].
	\label{eq:deltaECE_v2}
\end{equation}
Let us first consider the case of a superconducting (SC) transition, where we re-label $T_0\to T_c$. In the normal state ($T>T_c$), within a mean-field description, the contributions to the specific heat and to the ECE coefficient coming from the superconducting degrees of freedom vanish, i.e. $C_\epsilon(T_c^+),\text{ }\eta(T_c^+)=0$. This greatly simplifies Eq. (\ref{eq:deltaECE_v2}) to
\begin{equation}
	[\delta \eta]_{T_c} = \frac{dT_c}{d\epsilon}
	\label{eq:deltaECE_simple}
\end{equation}
Consider now that a second superconducting transition takes place inside the SC state at $T^*<T_c$. We will provide a specific model below. In this case, the SC order parameter is nonzero above $T^*$ and so $C_\epsilon(T^{*,+})$ and $\eta(T^{*,+})$ will be generically nonzero. Consequently, a simple form like Eq. (\ref{eq:deltaECE_simple}) does not hold for $T^*$ and one must resort to using the general form of Eq. (\ref{eq:deltaECE}) instead. Because of the $\eta(T^{*,-})$ term appearing on the right side of the equation, no general statements can be made about the relative sign between $[\delta\eta]_{T^*}$ and $dT^*/d\epsilon$ without appealing to a specific model. As such, we construct a Landau mean-field theory of coupled superconducting order parameters, stabilized by nematicity, using a symmetry-based approach, and subsequently minimize this free energy to obtain the SC order parameters and specific heat jumps as functions of the Landau parameters.

Specifically, we compute the temperature-dependent elastocaloric effect for a superconductor with nearby leading $s$-wave and subleading $d$-wave instabilities in the presence of nematic order. It is well-established that, within weak-coupling, the near-degeneracy of the superconducting order parameters can lead to time-reversal symmetry-breaking (TRSB) at a temperature $T^* <T_c$ even in the presence of nematic order \cite{KangPRB2018}. Our goal here is to compute the elastocaloric effect across both $T_c$ and $T^*$.  To do that, we construct a phenomenological Landau theory of coupled nematic $\Phi$, $s$-wave $\Delta_s = \mathrm{e}^{i \theta_s} \, |\Delta_s|$, and $d$-wave $\Delta_d = \mathrm{e}^{i \theta_d} \, |\Delta_d|$ superconducting order parameters. The terms of the free energy $\mathcal{F}[\Delta_s,\Delta_d,\Phi] = \mathcal{F}_{\Delta-\Phi} + \mathcal{F}_\Phi$ allowed by gauge, time-reversal, and point-group symmetries are (see e.g. \cite{FernandesPRL2013,Chen2020}):
\begin{equation}
	\begin{aligned}
		\mathcal{F}_{\Delta - \Phi} &= \frac{a_1}{2}|\Delta_s|^2+\frac{a_2}{2}|\Delta_d|^2-\lambda \Phi (\Delta_s^* \Delta_d + \Delta_d^* \Delta_s)+\frac{u_1}{4}|\Delta_s|^4+\frac{u_2}{4}|\Delta_d|^4 \\
		&+\frac{1}{2}(u_3 + u_4 \cos 2\phi)|\Delta_s|^2 |\Delta_d|^2
	\end{aligned}
	\label{eq:FE_unscaled}
\end{equation} 
where $\phi = \theta_s - \theta_d$ is the relative phase. Here, the phenomenological parameters $a_{1} = a_{1,0}\frac{T-T_1}{T_1}$ and $a_{2} = a_{2,0}\frac{T-T_2}{T_1}$, along with the trilinear coupling constant $\lambda$, determine $T_c$. The temperature scales $T_1$ and $T_2$ refer to the bare (i.e. decoupled) $s$- and $d$-wave superconducting (SC) transitions, respectively, in the tetragonal phase (where $\Phi=0$). We constrain the quartic coefficients to satisfy $u_1, u_2 >0$ and $u_1 u_2 > (u_3 +|u_4|)^2$, such that the free energy is bounded. Additionally, we focus our analysis in the regime well inside the nematic state, where $\Phi$ has saturated, such that the trilinear term may be effectively reduced to a bilinear coupling between $\Delta_s$ and $\Delta_d$ with effective scale $g_0 = \lambda \Phi_0$. For the same reason, we can also neglect $\mathcal{F}_\Phi$ and relabel $\mathcal{F}_{\Delta - \Phi} \to \mathcal{F}$. Hereafter, for concreteness, we set $g_0 > 0$.

The uniaxial strain applied in the experimental setup necessarily generates an $A_{1g}$ (i.e. symmetry-preserving) strain component $\epsilon_{A_{1g}} = \epsilon_{xx}+\epsilon_{yy}$, which is related to the applied uniaxial strain by the Poisson ratio. In what follows, we suppress the subscript $\epsilon_{A_{1g}}\to \epsilon$. Its main effect is to shift the SC transition temperature, which, to leading order, can be modeled as a  renormalization of the Landau coefficients of the terms that are quadratic in $\Delta_{s,d}$. Consequently, we introduce $\tilde{g}_0 = g_0 + g \epsilon$ and $\tilde{a}_{1,2} =a_{1,2}+h_{1,2}\epsilon$.

Since many phenomenological coefficients appear in $\mathcal{F}$, it is convenient to apply a dimensionless re-scaling in terms of three positive coefficients, $u_1$, $u_2$, and $g_0$. We thus define a set of primed quantities: $\Delta_s' = \frac{u_1^{3/8}u_2^{1/8}}{\sqrt{g_0}}\Delta_s$, $\Delta_d' = \frac{u_2^{3/8}u_1^{1/8}}{\sqrt{g_0}}\Delta_d$, $\tilde{a_1}' = \frac{(u_2/u_1)^{1/4}}{g_0}\tilde{a_1}$, $\tilde{a_2}' = \frac{(u_1/u_2)^{1/4}}{g_0}\tilde{a_2}$, $\mathcal{F}' = \frac{\sqrt{u_1 u_2}}{g_0^2}\mathcal{F}$, $u_{3,4}' = \frac{u_{3,4}}{\sqrt{u_1 u_2}}$, and $g' = g/g_0$. Likewise, from the expressions for $\tilde{a}_{1,2}'$ we also have $h_1' = \frac{(u_2/u_1)^{1/4}}{g_0}h_1$, $h_2' = \frac{(u_1/u_2)^{1/4}}{g_0}h_2$ and similarly $a_{1,0}' = \frac{(u_2/u_1)^{1/4}}{g_0} a_{1,0}$, $a_{2,0}' =\frac{(u_1/u_2)^{1/4}}{g_0} a_{2,0}$. Upon dispensing of the primes from everything except the temperature-dependent coefficients $\tilde{a}_{1,2}'$, the free energy becomes
\begin{equation}
	\begin{aligned}
		\mathcal{F} &= \frac{\tilde{a}_1'}{2}\Delta_s^2+\frac{\tilde{a}_2'}{2}\Delta_d^2-(1+g \epsilon)\Delta_s \Delta_d \cos \phi+\frac{1}{4}\Delta_s^4+\frac{1}{4}\Delta_d^4 +\frac{1}{2}(u_3 + u_4 \cos 2\phi)\Delta_s^2 \Delta_d^2
	\end{aligned}
	\label{eq:FE_scaled}
\end{equation} 
where, to simplify the notation, we have also removed the absolute value of the gap functions, such that $\Delta_{s,d}\ge 0$.

For later use, it is convenient to express the entropy $S$ and the elastocaloric (ECE) coefficient $\eta$ in terms of the SC order parameters. A total temperature derivative of $\mathcal{F}$ yields minus the entropy, but because $\Delta_{s,d}$ are assumed to satisfy $\partial \mathcal {F}/\partial \Delta_{s,d}=0$, it follows that only a partial temperature derivative of $\mathcal{F}$ is needed. Consequently, upon differentiating Eq. (\ref{eq:FE_unscaled}), $S$ may be expressed as a quadratic function of $\Delta_i$:
\begin{equation}
	S = -\frac{\partial \mathcal {F}}{\partial T} = -\frac{a_{1,0}}{2T_1}\Delta_s^2 - \frac{a_{2,0}}{2T_1}\Delta_d^2 
	\label{eq:entropy}
\end{equation}
We can then write the ECE coefficient in terms of $\Delta_i$: 

\begin{equation}
	\eta = - \frac{(\partial S/\partial \epsilon)_T}{(\partial S/\partial T)_\epsilon} = -\frac{a_{1,0} \Delta_s \frac{\partial \Delta_s}{\partial \epsilon}+a_{2,0} \Delta_d \frac{\partial \Delta_d}{\partial \epsilon}}{a_{1,0} \Delta_s \frac{\partial \Delta_s}{\partial T}+a_{2,0} \Delta_d \frac{\partial \Delta_d}{\partial T}}
	\label{eq:ECE}
\end{equation}

Note that we can equally use primed or non-primed variables. We now minimize Eq. (\ref{eq:FE_scaled}) and calculate $\Delta_i$ as explicit functions of $T$, $\epsilon$, and all other parameters. Indeed, the phase diagram of Eq. (\ref{eq:FE_unscaled}) is non-trivial when the bare $s$-wave and $d$-wave SC transitions are not too far way, as previously obtained in e.g. Refs. \cite{FernandesPRL2013,KangPRB2018}. Let us re-obtain these results to set the stage to compute the elastocaloric effect for all $T$ by employing Eq. (\ref{eq:ECE}). Upon lowering the temperature, the system first undergoes a second-order SC transition at $T_c$ into a time-reversal invariant $s+d$ state in which $\Delta_s$ and $\Delta_d$ condense simultaneously with a relative phase $\phi$ equal to zero. Upon analyzing the $\phi$-dependent part of $\mathcal{F}$

\begin{equation}
	\delta \mathcal{F} = \frac{1}{2}\Delta_s \Delta_d [-2(1+g \epsilon)\cos \phi + u_4 \Delta_s \Delta_d \cos 2\phi]
\end{equation} it is straightforward to see that $\delta \mathcal{F}$ is minimized by $\phi=0$  when 
\begin{equation}
	\Delta_s \Delta_d < \frac{1+g \epsilon}{2 u_4}
\end{equation} and by a non-zero $\phi$ when 
\begin{equation}
	\Delta_s \Delta_d > \frac{1+g \epsilon}{2 u_4}.
\end{equation}
The first case simply describes the $s+d$ state, and the associated condition is always satisfied at $T_c$ since $\Delta_{s,d}$ are infinitesimally small. However, as temperature is lowered and the order parameters increase, the latter condition can be satisfied at some $T^* < T_c$ signalled by $\Delta_s \Delta_d = (1+g \epsilon)/2u_4$. Of course, this threshold condition requires $T_1$ and $T_2$ to be comparable, otherwise the induced sub-leading order parameter $\Delta_d$ is too small \cite{FernandesPRL2013}. At $T^*$, the system undergoes another second-order transition into an exotic $s+\mathrm{e}^{i \phi}d$ state characterized by broken time-reversal symmetry with $\phi$ assuming a temperature-dependent value different from $0$ or $\pi$. 

We now proceed to determine $T_c$ and $T^*$ in terms of the Landau coefficients. Below $T_c$, where $\phi=0$, the free energy can be written as

\begin{equation}
	\label{eq:2}
	\mathcal{F}_\text{s+d}=\frac{\tilde{a}_1'}{2}\Delta_s^2 + \frac{\tilde{a}_2'}{2}\Delta_d^2 - (1+g \epsilon)\Delta_s \Delta_d+\frac{1}{4}\Delta_s^4 + \frac{1}{4}\Delta_d^4+\frac{u_3 + u_4}{2}\Delta_s^2 \Delta_d^2.
\end{equation}
Minimization with respect to the two components $\Delta_s$ and $\Delta_d$ yields a set of non-linear equations that may be solved perturbatively or numerically. To gain some insight about the behavior of the order parameter near $T_c$, it is sufficient to diagonalize the quadratic part of $\mathcal{F}_\text{s+d}$ by introducing ``rotated'' order parameters $\Delta_+$ and $\Delta_-$ defined by
\begin{equation}
	\begin{pmatrix}\Delta_s \\ \Delta_d\end{pmatrix} 
	= {\bf{v}}_+ \Delta_+ + {\bf{v}}_- \Delta_-.
	\label{eq:Delta_s+d_0}
\end{equation} Here, ${\bf{v}}_\pm$ are the orthonormal eigenvectors of the quadratic-form matrix 
\begin{equation}
	\begin{pmatrix} \tilde{a}_1' & -(1+g \epsilon) \\ -(1+g \epsilon) & \tilde{a}_2'\end{pmatrix}
\end{equation} and are given by 
\begin{equation}
	{\bf{v}}_\pm = \frac{\begin{pmatrix} -(\tilde{a}_1' - \tilde{a}_2') \mp \sqrt{(\tilde{a}_1' - \tilde{a_2}')^2+4(1+g \epsilon)} \\ 2(1+g \epsilon)\end{pmatrix}}{\sqrt{4(1+g \epsilon)^2 + [-(\tilde{a}_1' - \tilde{a}_2') \mp \sqrt{(\tilde{a}_1' - \tilde{a_2}')^2+4(1+g \epsilon)}]^2}}
\end{equation}
with corresponding eigenvalues 
\begin{equation}
	\lambda_\pm = \frac{\tilde{a}_1' + \tilde{a}_2' \pm \sqrt{(\tilde{a}_1' - \tilde{a}_2')^2+4(1+g \epsilon)^2}}{2}.
	\label{eq:eigenvalues}
\end{equation}
At $T_c$, the smaller eigenvalue $\lambda_-$ vanishes, signalling condensation of $\Delta_-$, whereas $\lambda_+$ remains positive. Nonetheless, there is no symmetry reason for why $\Delta_+$ should remain $0$ below $T_c$, as would be the case in a pure $\phi^4$-type model, because the basis transformation introduces higher-order terms like $\Delta_+ \Delta_-^3$ and $\Delta_- \Delta_+^3$ into $\mathcal{F}_\text{s+d}$. However, since these mixed terms only appear at order $\Delta^4$, we may approximate $\Delta_+$=0 near the phase transition, which is formally exact for $T\to T_c$ where $\Delta_-$ is suppressed.

Upon setting $\Delta_+=0$ in Eq. (\ref{eq:Delta_s+d_0}) we obtain $\Delta_{s,d}$ through $\Delta_s \approx v_-^{(1)}\Delta_-$ and $\Delta_d \approx v_-^{(2)} \Delta_-$. Here, $v_-^{(i)}$ is the $i$-th component of ${\bf{v}}_-$. Thus, near $T_c$,  $\mathcal{F}_\text{s+d}$ assumes the familiar form 
\begin{equation}
	\mathcal{F}_{s+d}\approx \frac{1}{2}\lambda_- \Delta_-^2 + \frac{1}{4}U \Delta_-^4
\end{equation} 
with $U = [v_-^{(1)}]^4+[v_-^{(2)}]^4+2(u_3 + u_4)[v_-^{(1)}]^2[v_-^{(2)}]^2$. Upon minimizing it, we obtain $\Delta_- \approx \sqrt{|\lambda_-|/U}$, giving $\Delta_{s,d}=v_-^{(1),(2)}\sqrt{|\lambda_-|/U}$. Of course, near $T_c$, we find the usual form
\begin{equation}
	\Delta_{s,d}(T,\epsilon)\approx \Upsilon_{s,d}(\epsilon)\sqrt{T_c(\epsilon)-T} 
	\label{eq:Delta_s+d}
\end{equation} 
where $\Upsilon_{s,d}$ is a non-universal, strain-dependent pre-factor arising from taking the $T\to T_c$ limit in the full expressions for $\Delta_i$ involving $v^{(i)}$, $\lambda_-$ and $U$. We solve for $T_c$ by setting $\lambda_-=0$ and obtain
\begin{equation}
	\begin{aligned}
		T_c(\epsilon)&=\frac{T_1+T_2-T_1(h_1/a_{1,0}+h_2/a_{2,0})\epsilon}{2}\\ &+ \sqrt{\Bigg[\frac{T_1+T_2-T_1(h_1/a_{1,0}-h_2/a_{2,0})\epsilon}{2} \Bigg]^2 + T_1^2\frac{(1+g \epsilon)^2}{a_{1,0}a_{2,0}}} 
	\end{aligned}
	\label{eq:Tc}
\end{equation}
Note that there is no dependence on the quartic coefficients $u_3$ and $u_4$, since destabilization of the normal state is described by the quadratic part of $\mathcal {F}$ only.

Clearly, $\Delta_{s,d}=0$ above $T_c$, and hence both the specific heat and ECE coefficient above $T_c$ are expected to be zero, i.e. there should be no contribution to these thermodynamic quantities from superconducting degrees of freedom (within our mean-field description, of course). As a result, we obtain the expected result

\begin{equation}
	[\delta\eta]_{T_c} = \eta(T_c^-) = \frac{dT_c}{d\epsilon}
	\label{eq:ECE_Tc}
\end{equation}

\begin{figure}
	\centering
	\includegraphics[width=0.95\linewidth]{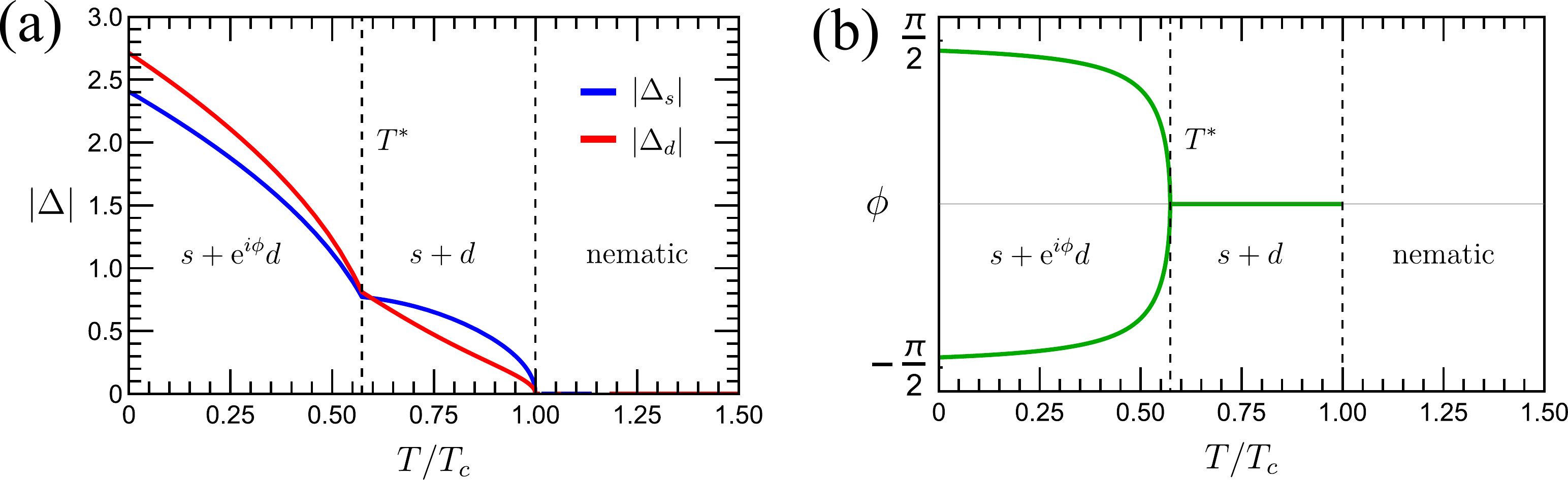}
	\caption{\label{fig:OP_vs_T} \textbf{Order parameters magnitude and relative phase as functions of temperature.} (a) The magnitudes of the s-wave (blue) and d-wave (red) superconducting order parameters obtained by minimizing Eq. (\ref{eq:FE_scaled}). Both the SC ($T_c$) and TRSB ($T^*$) transitions are visible, with $|\Delta_i|$ clearly changing slope at $T^*$. (b) The relative phase $\phi = \theta_s - \theta_b$  is formally undefined for $T>T_c$, is equal to zero for $T^* < T < T_c$ , and breaks time-reversal symmetry (the $\mathbb{Z}_2$ symmetry described by $\phi\to-\phi$) for $T<T^*$. For both plots $\epsilon=0$, $a_{1,0}=1$, $a_{2,0}=4$, $h_1=1$, $h_2=-10$, $g=1$, $u_3=0.15$, $u_4=0.8$, and $T_2/T_1=0.9$.} 
\end{figure}

The key difference between the normal-to-SC transition and the TRSB transition is that $\Delta_i$ are nonzero above the latter transition, as seen in \autoref{fig:OP_vs_T}(a) for $T\ge T^*$. Therefore,  $\eta$ and $C_\epsilon$ are nonzero above $T^*$. This means that the ECE coefficient discontinuity at $T^*$ does not follow the simple form of Eq. (\ref{eq:ECE_Tc}), and instead includes contributions from all terms in the $\epsilon$-$T$ Ehrenfest relation. 

To proceed,  we now investigate more closely the TRSB $s+d \to s+\mathrm{e}^{i\phi} d$ transition, where the relative phase $\phi$ unlocks from zero. Solving $\partial \mathcal{F}/\partial \phi =0$ gives 
\begin{equation}
	\cos \phi = \frac{1+g \epsilon}{2u_4 \Delta_s \Delta_d}
\end{equation}
from which the free energy in the $s+\mathrm{e}^{i\phi} d$ phase can be written without reference to the phase: 
\begin{equation}
	\mathcal{F}_{s+\mathrm{e}^{i\phi} d} = \frac{\tilde{a}_1'}{2}\Delta_s^2 + \frac{\tilde{a}_2'}{2}\Delta_d^2+\frac{1}{4}\Delta_s^4 + \frac{1}{4}\Delta_d^4 + \frac{u_3 - u_4}{2}\Delta_s^2 \Delta_d^2 - \frac{(1+g \epsilon)^2}{2u_4}.
\end{equation}
Minimization of $\mathcal{F}_{s+\mathrm{e}^{i\phi} d}$ yields the mean-field equations
\begin{align}
	\tilde{a}_1' + \Delta_s^2 + (u_3 - u_4)\Delta_d^2 = 0\\
	\tilde{a}_2' + \Delta_d^2 + (u_3 - u_4)\Delta_s^2 = 0
\end{align}
whose solution gives the gaps:

\begin{equation}
	\begin{aligned}
		\Delta_s = \sqrt{\frac{\tilde{a}_1' + (u_4 - u_3)\tilde{a}_2'}{(u_3 - u_4)^2-1}}\\
		\Delta_d = \sqrt{\frac{\tilde{a}_2' + (u_4 - u_3)\tilde{a}_1'}{(u_3 - u_4)^2-1}}
	\end{aligned}
	\label{eq:Delta_TRSB}
\end{equation}
and the relative phase:
\begin{equation}
	\phi = \arccos\Bigg[\frac{(1+g \epsilon)|(u_4-u_3)^2-1|}{2u_4\sqrt{(\tilde{a}_1' + (u_4 - u_3)\tilde{a}_2')(\tilde{a}_2' + (u_4 - u_3)\tilde{a}_1')}} \Bigg]
\end{equation}
The condition $\Delta_s \Delta_d = (1+g \epsilon)/2u_4$ (or equivalently $\phi=0$) on the above expressions yields an equation for $T^*$, which we solve to obtain
\begin{equation}
	T^*(\epsilon) = \frac{\tau_1+\tau_2}{2}-\sqrt{\Big(\frac{\tau_1-\tau_2}{2} \Big)^2+\Big(\frac{1+g\epsilon}{2u_4}\Big)^2 \frac{(\delta{u}^2-1)^2}{(a_{1,0}+\delta{u}\, a_{2,0})(a_{2,0}+\delta{u}\,a_{1,0})}}
	\label{eq:T*}
\end{equation}
where $\delta{u} = u_4 - u_3$ and 
\begin{equation}
	\begin{aligned}
		\tau_1 &= \frac{T_1 a_{1,0}+T_2 \,\delta{u}\,a_{2,0}-T_1(h_1 + \delta{u}\,h_2)\epsilon}{a_{1,0}+\delta{u}\,a_{2,0}}\\
		\tau_2 &= \frac{T_1 \, a_{2,0}+T_2 \delta{u}\,a_{1,0}-T_1(h_2 + \delta{u}\,h_1)\epsilon}{a_{2,0}+\delta{u}\,a_{1,0}}
	\end{aligned}
\end{equation}
The temperature evolution of the relative phase obtained from the minimization of the free energy is shown in \autoref{fig:OP_vs_T}(b). Note that $\phi$ approaches $\pm \pi/2$ as the temperature is decreased.

Recalling that $[\delta\eta]_{T^*} = \eta(T^{*,-})-\eta(T^{*,+})$ depends both on $dT^*/d\epsilon$ and $\eta(T^{*,-})$ via Eq. (\ref{eq:deltaECE}), and that $\delta C > 0$, we first calculate $\eta(T^{*,-})$. Using the expressions for $\Delta_s$ and $\Delta_d$ in the $s+\mathrm{e}^{i\phi} d$ phase from Eq. (\ref{eq:Delta_TRSB}), we see the $\Delta_i^2$ are functions of the form $\alpha_1 + \alpha_2 T + \alpha_3 \epsilon$ where $\alpha_1$, $\alpha_2$, and $\alpha_3$ are constants (generically different for the two pairing channels) which are independent of $T$ and $\epsilon$. From here, Eq. (\ref{eq:entropy}) implies that the entropy in this phase is a function of the form 
\begin{equation}
	S = \beta_1 + \beta_2 T + \beta_3 \epsilon
\end{equation}
where $\beta_1$, $\beta_2$, and $\beta_3$ are also constants independent of $T$ and $\epsilon$. Consequently, both $\partial S/\partial \epsilon$ and $\partial S/\partial T$ are independent of $T$ and $\epsilon$, leading to an ECE coefficient that is a constant in the $s+\mathrm{e}^{i\phi}d$ phase. Specifically, for any $T<T^*(\epsilon)$, we find

\begin{equation}
	\eta(T<T^{*}) = -T_1 \frac{a_{1,0} h_1 + a_{2,0} h_2 + \delta{u}(a_{1,0} h_2 + a_{2,0} h_1)}{a_{1,0}^2 + a_{2,0}^2+2\delta{u}\, a_{1,0} a_{2,0}}
	\label{eq:ECE_TRSB}
\end{equation}

Having derived analytical expressions for $T_c(\epsilon)$, Eq. (\ref{eq:Tc}); for  $T^*(\epsilon)$, Eq. (\ref{eq:T*}); and for $\eta(T^{*,-})$, Eq. (\ref{eq:ECE_TRSB}), it is of interest to determine if a particular range of parameters is compatible with a mismatch between the sign of $dT^*/d\epsilon$ and the sign of $[\delta\eta]_{T^*} \propto (dT^*/d\epsilon - \eta(T^{*,-}))$; recall that $\delta C >0$. More specifically, to be consistent with the experimental results in the underdoped sample, we are interested in a set of parameters for which $dT_c/d\epsilon = [\delta\eta]_{T_c} > 0$, $dT^*/d\epsilon>0$, and $[\delta\eta]_{T^*}<0$. Due to the presence of various phenomenological coefficients in both expressions, we provide a qualitative analysis first.

The signs of $dT_c/d\epsilon$ and $dT^*/d\epsilon$ will be governed by the three coefficients $h_1$, $h_2$, and $g$, which determine how the coefficients of the quadratic terms in the free-energy Eq. (\ref{eq:FE_scaled}) change as a function of strain. From Eq. (\ref{eq:Tc}), it is clear that $h_1, h_2 <0$  and $g>0$  lead to a $T_c$ that increases for increasing strain, $dT_c/d\epsilon > 0$, whereas $h_1, h_2 >0$  and $g<0$  lead to a  $T_c$ that decreases for increasing strain, $dT_c/d\epsilon < 0$. For other combinations of the signs of the three coefficients, we can only make a qualitative assessment of the resulting sign of $dT_c/d\epsilon$ if we also know the relative magnitudes of the coefficients.

As for $T^{*}$, setting for a moment $g = 0$ and focusing only on the effect of $h_{1,2}$, it is clear that the sign of $dT^*/d\epsilon$  will generally be the same as $dT_c/d\epsilon$. This is because the threshold condition for the emergence of the $s+\mathrm{e}^{i\phi}d$ phase, namely $\Delta_s \Delta_d \ge \frac{1}{2u_4}$, is met at a higher (lower) temperature if $T_c$ is higher (lower), since $\Delta_{s,d}$ increases (decreases) with increasing (decreasing) $T_c$, see e.g. Eq. (\ref{eq:Delta_s+d}). The same conclusion also follows from inspecting Eq. (\ref{eq:T*}) for $T^*$.

On the other hand, the effect of $g$ on $T^*$ is the opposite of the effect of $g$ on $T_c$, since a positive $g$ implies an enhancement of the threshold condition $\Delta_s \Delta_d = (1+g \epsilon)/2u_4$, making it harder for the system to order into the $s+\mathrm{e}^{i\phi}d$ state. Indeed, Eq. (\ref{eq:T*}) shows that $g>0$ favors $dT^*/d\epsilon<0$ whereas $g<0$ favors $dT^*/d\epsilon>0$.

\begin{figure}[h]
	\centering
	\includegraphics[width=0.65\linewidth]{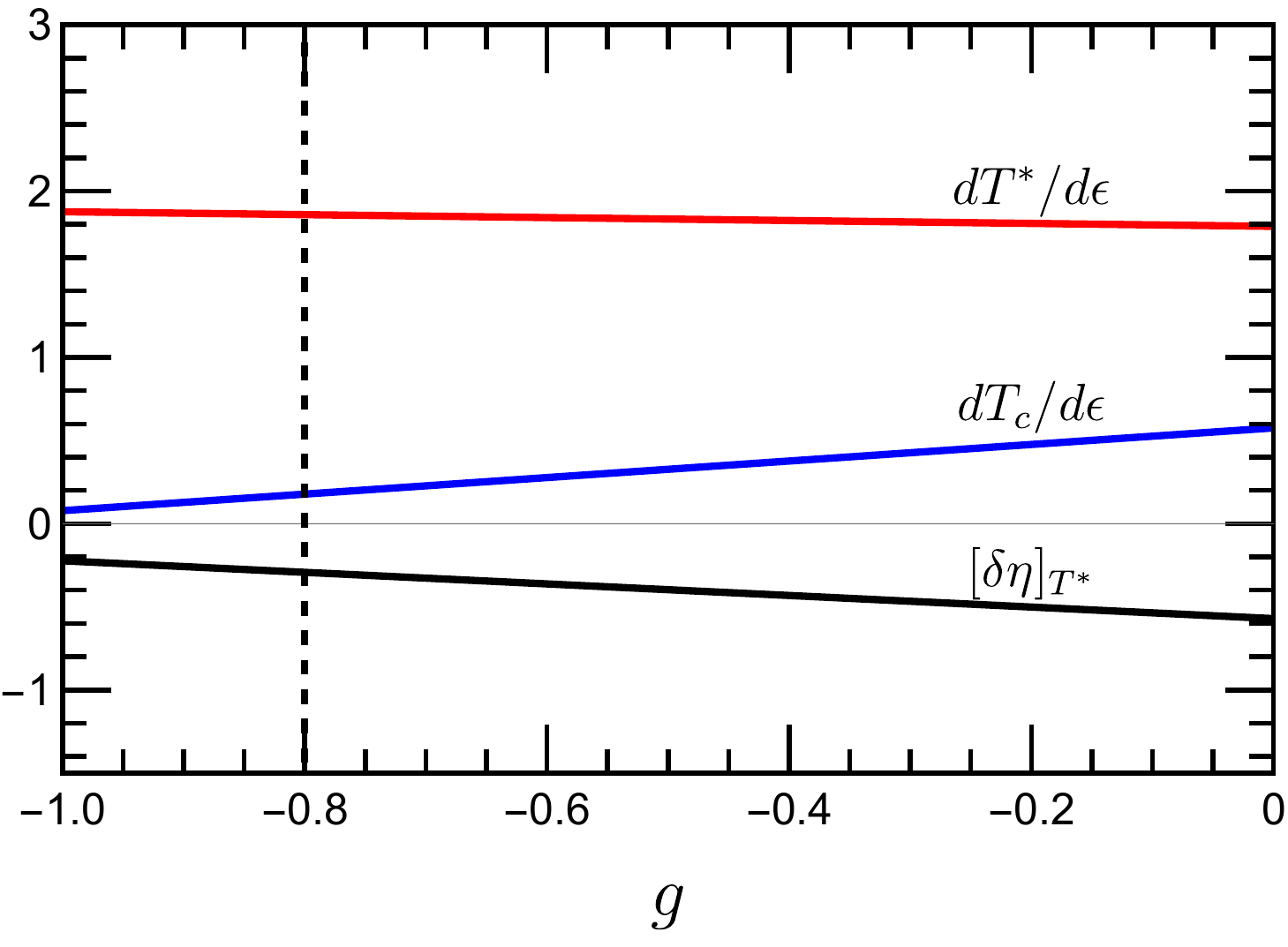}
	\caption{\label{fig:Tcrit_deriv_vs_g1}\textbf{$dT^*/d\epsilon$, $dT_c/d\epsilon$, and $[\delta\eta]_{T^*}$ as functions of $g$}. The three quantities are evaluated at $\epsilon=0$  and plotted as functions of $g$ using Eq. (\ref{eq:T*}), Eq. (\ref{eq:Tc}), and Eq. (\ref{eq:deltaECE}), respectively. The vertical axis is in units of $T_1$. Near $g\approx -0.8$ (dashed line), we recover the experimental result from Figure 3(b) of the main text that $dT^*/d\epsilon \approx 10 dT_c/d\epsilon>0$ with $[\delta\eta]_{T^*}<0$. The corresponding Landau parameters are the same as in \autoref{fig:OP_vs_T} (other than $g$ since it varies), namely $a_{1,0}=1$, $a_{2,0}=4$, $h_1=1$, $h_2 = -10$, $u_3=0.15$, $u_4=0.8$, $T_2/T_1=0.9$.}
\end{figure}

A qualitative comparison with the experimental results for $T_c (\epsilon)$ and $T^*(\epsilon)$ in the overdoped and underdoped compounds allows us to further constrain the range of values of $h_1$, $h_2$, and $g$ (assuming that they do not vary strongly with doping). Main Text Figure 1(a) shows that the overdoped $T_c$ decreases with increasing strain, suggesting that at least one of the $h_i$ must be positive. Since the state realized in overdoped compositions is purely $s$-wave, we conclude that $h_1>0$ . Now, from Figure 2(d), the underdoped $T_c$ increases with $\epsilon$, which suggests that $h_2$ should be negative and the dominant coefficient, i.e. $h_2 < 0$  and $|h_2|\gg h_1 >0$. . The same conditions generally favor $T^*$ increasing with strain in the underdoped composition. Indeed, expanding $T_c$ and $T^*$ in Eq. (\ref{eq:Tc}) and Eq. (\ref{eq:T*}) to linear order in strain in the regime $|h_2|\gg |h_1|$ gives:
\begin{equation}
	\begin{aligned}
		T_c(\epsilon) &\approx T_c(0) + (-\gamma_1 h_2 + \gamma_2 g)\epsilon\\
		T^*(\epsilon) &\approx T^*(0) + (-\gamma_3 h_2 - \gamma_4 g )\epsilon
	\end{aligned}
\end{equation}
where, crucially, $\gamma_i$ are all positive constants. To constrain the possible values of $g$, we can use the experimental result of Figure 3(b) of the main text that  $dT^*/d\epsilon \approx 10 dT_c/d\epsilon$. 
\begin{figure}
	\centering
	\includegraphics[width=0.65\linewidth]{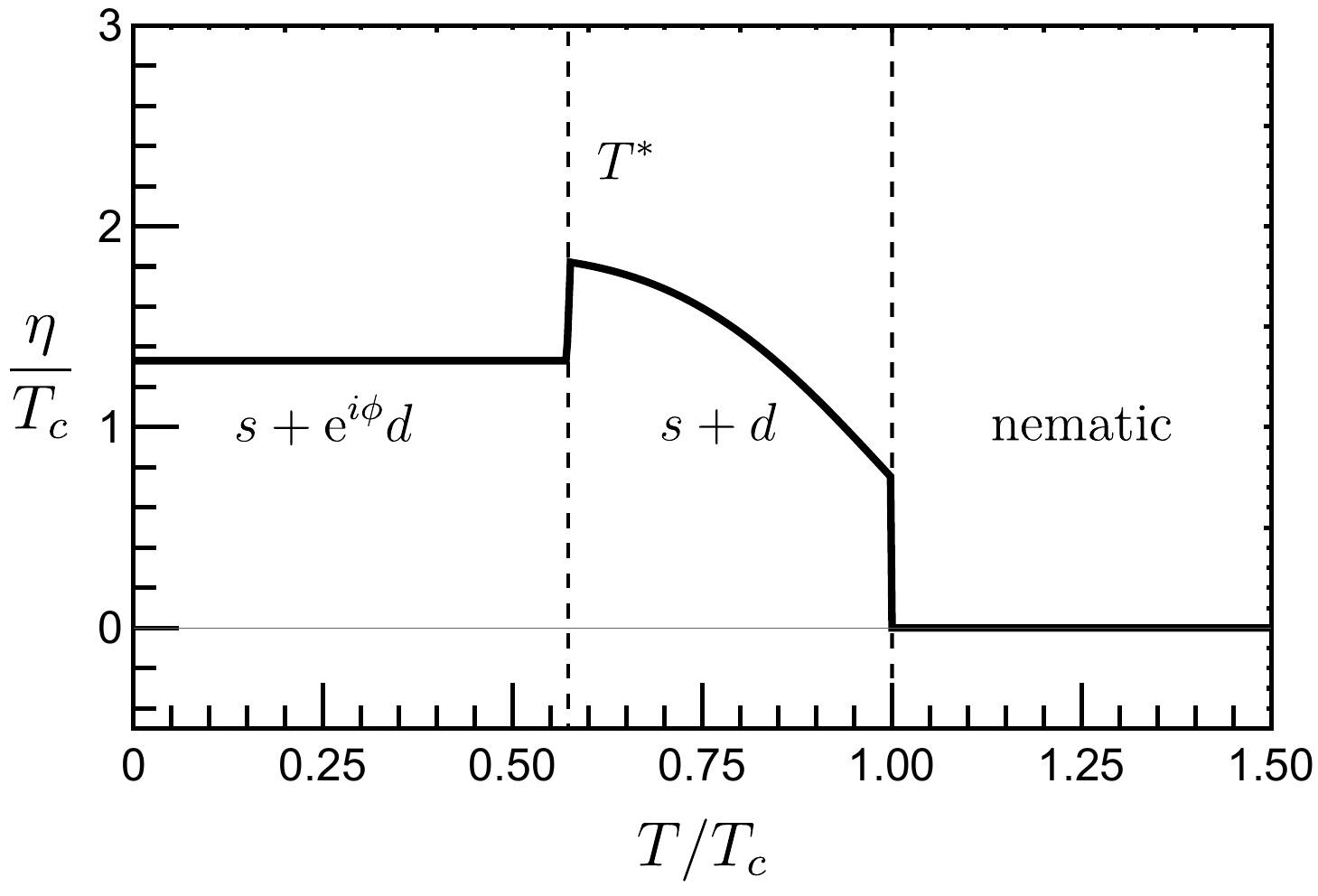}
	\caption{\label{fig:ECE_vs_eps} \textbf{Elastocaloric coefficient as a function of temperature.} ECE coefficient $\eta$ obtained from minimizing the Landau free energy, which shows discontinuities with opposite signs at $T=T^*$ and $T=T_c$. The corresponding Landau parameters are the same as in \autoref{fig:OP_vs_T}, namely, $a_{1,0}=1$, $a_{2,0}=4$, $h_1=1$, $h_2 = -10$, $g=1$, $u_3=0.15$, $u_4=0.8$, $T_2/T_1=0.9$. Note that the signs of the ECE discontinuities shown here are opposite to the experimentally measured ones (in Main Text Figure 2 and 3, for example). This arises due to positive strain on the sample leading to a negative temperature change, i.e., cooling (see \autoref{fig:fscan}) and we plot the absolute temperature change in the experimental plots.}
\end{figure}

The final experimental constraint is that $dT^*/d\epsilon < \eta(T^{*,-})$, such that the ECE coefficient jump at $T^*$ has the opposite sign as $dT^*/d\epsilon$, i.e. $[\delta\eta]_{T^*}<0$. The crucial point from Eq. (\ref{eq:ECE_TRSB}) is that $\eta(T^{*,-})$ does not depend on $g$ or $h_i$, whereas $dT^*/d\epsilon$ does. Therefore, there should be a wide range of Landau coefficients for which the two conditions $dT^*/d\epsilon \gg dT_c/d\epsilon > 0$ and $dT^*/d\epsilon < \eta(T^{*,-})$ are satisfied. To illustrate that, in \autoref{fig:Tcrit_deriv_vs_g1} we plot $[\delta\eta]_{T^*}$, $dT^*/d\epsilon$, and $dT_c/d\epsilon$ in the zero strain limit as functions of only the off-diagonal Landau parameter $g$ and with all other Landau parameters fixed ($h_2<0$ and $|h_2|=10h_1>0$; the other parameters are given in the figure caption). The dashed line, corresponding to $g \approx -0.8$, gives the desired ratio $dT^*/d\epsilon \approx 10 dT_c/d\epsilon$ as well as $[\delta\eta]_{T^*} < 0$. In \autoref{fig:ECE_vs_eps}, we show the full temperature dependence of $\eta(T)$ for these Landau coefficients, obtained by numerically minimizing the Landau free energy and using Eq. (\ref{eq:ECE}).

Of course, there are many other parameters that give similar results, even if we only change the ratio $|h_2|/h_1$. The point of this exercise is to show that the model for $s+\mathrm{e}^{i\phi}d$ superconductivity introduced here is perfectly compatible with the key experimental result that $dT^*/d\epsilon$ and $[\delta\eta]_{T^*}$ have opposite signs. In contrast, if the transition at $T^*$ was due to other (i.e. non-superconducting) degrees of freedom, one would expect the generic relationship in Eq. (\ref{eq:deltaECE_simple}) to hold, $[\delta\eta]_{T^*} = dT^*/d\epsilon$.

\subsection*{Consistency check from ECE Ehrenfest relation}


We use the Ehrenfest relation derived above (Eq. (\ref{eq:ehrenECE})) to relate the ECE discontinuity to the specific heat discontinuity $\Delta C_p$ and strain derivative of the transition temperature, $dT_{0}/d\epsilon$,
\begin{equation}
	\frac{dT_{0}}{d\epsilon}=\frac{\delta\left[C_p\frac{\partial T}{\partial \epsilon}\right]}{\delta C_p}.
	\label{eq:ehren}
\end{equation}.
To verify this for our experimental ECE data, we first calculate the quantity $\delta\left[C_p\frac{\partial T}{\partial \epsilon}\right]/\delta C_p$ using the specific heat (at zero strain) for a sample of similar doping (see \autoref{fig:char}(a)). From our experimentally measured ECE ($\partial T/\partial\epsilon$, with $\partial\epsilon=5\times10^{-5}$) at various DC strains, we can calculate this quantity as a function of strain. It appears to follow the strain dependence of the ECE discontinuity $\delta\left[\partial T/\partial \epsilon\right]$ at the transition. The ECE measurements also tracks the phase transition at different bias strains, from which we can extract $T_{crit}(\epsilon)$ precisely. From Eq. (\ref{eq:ehren}), integrating the calculated $\delta\left[C_p\frac{\partial T}{\partial \epsilon}\right]/\delta C_p$ as a function of strain is expected to reproduce the experimental $T_{0}(\epsilon)$ up to a constant. This constant is the zero-strain $T_{0}$. An important technical consideration here comes from our thermometry being not ideal, that is, we do not measure the entire temperature change induced in the sample by strain. Thus we need to multiply $\partial T/\partial\epsilon$ by a ``non-adiabaticity" constant to exactly match the measured $T_{0}(\epsilon)$. This indicates what fraction of temperature change induced in the sample is actually measured at the thermocouple. For our measurements, we find the non-adiabaticity constant to be about 3, implying that the thermocouple measures one-third of the temperature change induced in the sample.

\begin{figure*}[h!]
	\centering
	\includegraphics[width=0.99\linewidth]{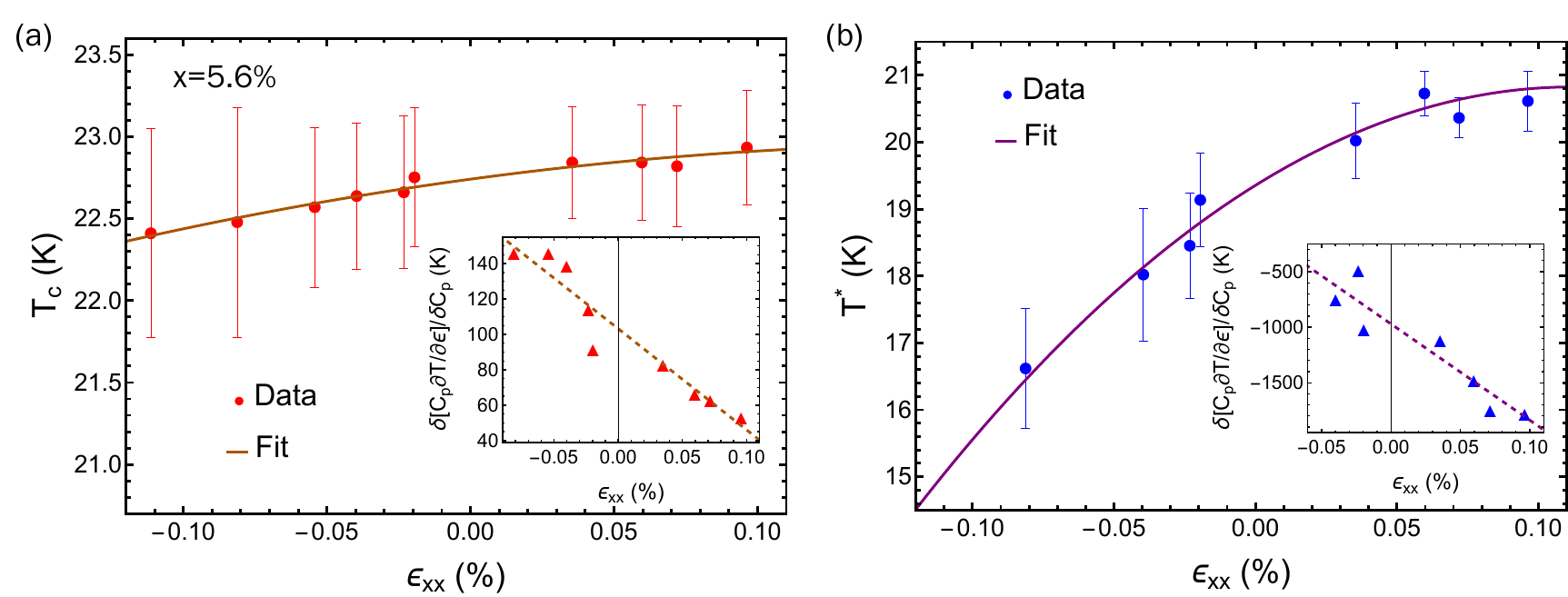}
	\caption{\label{fig:ehren} \textbf{Fit to \Tc and $T^*$ using Ehrenfest relation.} The transition temperatures obtained from the ECE data, (a) \Tc and (b) $T^*$, for a slightly underdoped ($x=5.6\%$) sample are plotted. Insets show the magnitude of the ECE discontinuities at the corresponding transitions, scaled by the specific heat discontinuities according to the Ehrenfest relation Eq. (\ref{eq:ehren}). The quadratic fits to \Tc and $T^*$ are obtained from integrating the linear fits to the data in insets (see text for details). Note the significantly reduced vertical scale in (a) as compared to (b).} 
\end{figure*}

For our overdoped sample, the ECE discontinuity at \Tc is approximately constant for different strains (Main Text Figure 1), implying a constant $\delta\left[C_p\frac{\partial T}{\partial \epsilon}\right]/\delta C_p$ as a function of strain. Integrating this yields a linear dependence of \Tc on strain $\epsilon_{xx}$, which is also observed experimentally.

For the underdoped samples, the ECE discontinuities at both \Tc and $T^*$ are seen to vary linearly with strain (see insets in \autoref{fig:ehren}). This, after integrating, leads to a quadratic dependence of both \Tc and $T^*$ on strain. This is non-trivial compared to the overdoped data since a quadratic fit has more coefficients than a linear fit, with these coefficients determined from the fit to how the ECE discontinuities vary with strain. The fact that the calculated $T_c(\epsilon)$ and $T^*(\epsilon)$ matches the experimentally measured ones establishes that both are well-behaved thermodynamic transitions. We use the same non-adiabaticity constant to fit both \Tc and $T^*$, plotted in \autoref{fig:ehren} (these are the same fits shown in Main text Figure 3). Since the $T^*$ transition is not seen in specific heat, we use a value of $\delta C_p|_{T^*}$ that best matches the data. We find that $\delta C_p|_{T^*}$ is only 3$\%$ of the specific heat discontinuity at $T_c$ ($\delta C_p|_{T_c}$), explaining why this transition is not seen in specific heat measurements.

For the quadratic fit to $T^*(\epsilon)$, we find that the linear coefficient has the opposite sign compared to what would be expected from how the ECE discontinuity at $T^*$ changes with strain. In other words, the sign of the ECE discontinuity at $T^*$ is opposite to $dT^*/d\epsilon$. As discussed above in details in the section ``Theoretical calculation of the elastocaloric effect in an $s+\mathrm{e}^{i\phi} d$ superconductor", a time-reversal symmetry breaking transition within a superconducting state may naturally account for this.

\subsection*{Relation to thermal expansion measurements}

Thermal expansion is a thermodynamic quantity which is related to strain derivatives of the transition temperatures. In particular, the discontinuity in the thermal expansion coefficient at a phase transition is related to $dT_{0}/d\epsilon$ through the Ehrenfest relation \cite{TestardiPRB1976},
\begin{equation}
	\delta\alpha_a=\frac{\delta C_p}{T_{0}}\bigg(\frac{dT_{0}}{d\sigma_{xx}}\bigg)=\frac{\delta C_p}{T_{0}}\frac{1}{Y_a}\bigg(\frac{dT_{0}}{d\epsilon_{xx}}\bigg),
\end{equation}
where $\sigma_{xx}$ is uniaxial stress and $Y_a$ is the Young's modulus corresponding to $a$-axis stress. We can therefore use the magnitude of the expected specific heat discontinuity at $T^*$ ($\delta C_p|_{T^*}$), and our experimentally determined $dT_c/d\epsilon$ and $dT^*/d\epsilon$, to estimate the discontinuities that should be seen in thermal expansion coefficients at \Tc and $T^*$. We estimate $\frac{\delta\alpha_a|_{T_c}}{\delta\alpha_a|_{T^*}}\approx4$.

We compare this to the measurements reported in Ref. \cite{MeingastPRL2012}, where we believe the 5.5$\%$ sample shows two anomalies around \Tc for $a$-axis thermal expansion ($\alpha_a$). Only one anomaly is seen in $c$-axis thermal expansion ($\alpha_c$), and no second anomaly is seen in either 4.5$\%$ or 6.5$\%$ samples, which are the nearest dopings around 5.5$\%$ they measured. From the magnitudes of the two anomalies in $\alpha_a$ measurements, we find $\frac{\delta\alpha_a|_{T_c}}{\delta\alpha_a|_{T^*}}\approx4$, which agrees extremely well with our estimation based on the ECE data.


\begin{thebibliography}{49}
\bibitem[irr()]{irrep}
	Any arbitrary strain in a crystalline solid can be decomposed into strains that
	have the same symmetry properties as the irreducible representations (irreps)
	of the crystal's point group. $\mathrm{W}$e refer to these as the irreducible
	strains. $\mathrm{I}$n a tetragonal lattice, the irreducible strains are: two
	compressions, in-plane and out-of-plane, transforming as the
	$\mathrm{A_{1g}}$ irrep, and three shear strains, transforming as the
	$\mathrm{B_{1g}}$, $\mathrm{B_{2g}}$ and $\mathrm{E_{g}}$ irreps.
\bibitem[RUS()]{RUSjump}
	We note that although we have identified the $c_{A_{1g}}$ in the above equation
	with the in-plane \aog modulus $(c_{11}+c_{12})/2$, there is an out-of-plane
	\aog modulus $c_{33}$ which can also show a jump at $\mathrm{T}_c$.
	$\mathrm{H}$owever, d${T}_c/d\epsilon$ for out-of-plane strain is expected to
	be a factor of 10 larger than for in-plane strain, and thus predicts a factor
	of 100 larger jump in $c_{A_{1g}}$ for out-of-plane strain. $\mathrm{D}$ue to
	this two orders of magnitude difference between $c_{A_{1g}}$ discontinuities
	for in-plane and out-of-plane \aog strains, we can safely associate the
	measured jump with the in-plane modulus $(c_{11}+c_{12})/2$.
\end{thebibliography}
\end{document}